\documentclass[12pt]{article}
\usepackage[round, authoryear]{natbib}
\usepackage{mathrsfs}
\usepackage[psamsfonts]{amssymb}
\usepackage{natbib,graphicx,setspace,lscape,longtable}
\usepackage{xcolor}
\usepackage{epsfig}
\usepackage{multirow}
\usepackage{mathrsfs,mathtools,amsmath,amsthm,color,url}
\usepackage{verbatim}
\usepackage[shortlabels]{enumitem}
\usepackage[margin=1in]{geometry}
\renewcommand{\baselinestretch}{1.15}
\usepackage{bm}
\makeatletter
\def\singlespace{\def\baselinestretch{1}\@normalsize}

\usepackage[colorlinks=true,allcolors=blue]{hyperref}%
\usepackage{breakurl}

\UrlBigBreaks

\bibpunct{(}{)}{;}{a}{,}{,}
\newtheorem{theorem}{Theorem}
\newtheorem{lemma}{Lemma}
\newtheorem{proposition}{Proposition}

\newtheorem{corollary}{Corollary} 

\theoremstyle{definition}

\DeclarePairedDelimiter{\ceil}{\lceil}{\rceil}

\def\beq{\begin{equation}}
\def\eeq{\end{equation}}
\def\beqr{\begin{eqnarray}}
\def\eeqr{\end{eqnarray}}
\def\beqrs{\begin{eqnarray*}}
\def\eeqrs{\end{eqnarray*}}
\def\bet{\begin{theorem}}
\def\eet{\end{theorem}}
\def\bel{\begin{lemma}}
\def\eel{\end{lemma}}
\def\bep{\begin{proposition}}
\def\eep{\end{proposition}}
\def\bg{\begin{figure}[tbph]\begin{center}}
\def\eg{\end{center}\end{figure}}

\def\bc{\begin{center}}
\def\ec{\end{center}}

\renewcommand{\thefootnote}{\fnsymbol{footnote}}
\numberwithin{equation}{section}

\AtBeginDocument{%
   \setlength\abovedisplayskip{7.5pt}
   \setlength\belowdisplayskip{7.5pt}
}

\begin{document}

\title{Using Maximum Entry-Wise Deviation to Test the Goodness-of-Fit for Stochastic Block Models}
\author{{Jianwei Hu{$^*$}, Jingfei Zhang{$^*$}, Hong Qin, Ting Yan and Ji Zhu}}
\date{}
\maketitle{}

\begin{singlespace}
\begin{footnotetext}[1]
{The first two authors contributed equally to this work.
Jianwei Hu, Department of Statistics, Central China Normal University, Wuhan, P.R. China, 430079, Email: jwhu@mail.ccnu.edu.cn.
Jingfei Zhang, Department of Management Science, University of Miami, Coral Gables, FL 33124, Email: ezhang@bus.miami.edu.
Hong Qin, Department of Statistics, Central China Normal University, Wuhan, P.R. China, 430079, Email: qinhong@mail.ccnu.edu.cn.
Ting Yan, Department of Statistics, Central China Normal University, Wuhan, P.R. China, 430079, Email: tingyanty@mail.ccnu.edu.cn.
Ji Zhu, Department of statistics, University of Michigan, Ann Arbor, MI 48109, Email: jizhu@umich.edu.
}
\end{footnotetext}
\end{singlespace}
\vspace{-0.2in}
\begin{abstract}
The stochastic block model is widely used for detecting community structures in network data.
How to test the goodness-of-fit of the model is one of the fundamental problems and
has gained growing interests in recent years.
In this article, we propose a novel goodness-of-fit test based on the maximum entry of the centered
and re-scaled
adjacency matrix
for the stochastic block model. One noticeable advantage of the proposed test is that the number of communities can be allowed to grow linearly with the number of nodes ignoring a logarithmic factor.  We prove that the null distribution of the test statistic converges in distribution to a Gumbel distribution, and we show that both the number of communities and the membership vector can be tested via the proposed method.
Further, we show that the proposed test has asymptotic power guarantee against a class of alternatives.
We also demonstrate that the proposed method can be extended to the degree-corrected stochastic block model.
Both simulation studies and real-world data examples indicate that the proposed method works well.\\


\end{abstract}

\newpage
\section{Introduction}

One of the fundamental problems in network data analysis is community detection that aims to divide nodes into communities
such that the links are dense within communities and relatively sparse between communities.
The stochastic block model proposed by \cite{Holland:Laskey:Leinhardt:1983} is probably the most studied network model for this purpose;
see \cite{Snijders:Nowicki:1997}, \cite{Nowicki:Snijders:2001}, \cite{Bickel:Chen:2009}, \cite{Rohe:Chatterjee:Yu:2011}, \cite{Choi:Wolfe:Airoldi:2012}, \cite{Jin:2015} and \cite{Zhang:Zhou:2016} for some of the representative work.

In a stochastic block model with $k$ communities, $n$ nodes are clustered into $k$ blocks,
i.e., there exists a mapping $\sigma$ of community membership:  $[n]\rightarrow[k]^n$, where $[n]=\{1,\ldots,n\}$.
Given the community membership $\sigma$, the entries $A_{ij}$ ($i>j$) of the symmetric adjacency matrix $A\in \{0, 1\}^{n\times n}$ of an undirected random graph $\mathcal{G}$ are then assumed to be mutually independent Bernoulli random variables with the occurrence probabilities
$P_{ij}=B_{\sigma(i)\sigma(j)}$ for certain symmetric probability matrix $B\in [0,1]^{k\times k}$.
A large number of methods for recovering the community membership have been proposed,
including modularity \citep{Newman:2006}, profile-likelihood maximization \citep{Bickel:Chen:2009}, pseudo-likelihood maximization
\citep{Amini:2013}, variational methods \citep{Daudin:Picard:Robin:2008} and spectral clustering
\citep{Rohe:Chatterjee:Yu:2011, Jin:2015}.
Asymptotic properties of the estimators of the community membership have also been established; see \cite{Choi:Wolfe:Airoldi:2012},
\cite{Rohe:Chatterjee:Yu:2011}, \cite{Zhao:Levina:Zhu:2012}, \cite{Sarkar:Bickel:2015}, \cite{Jin:2015}, \cite{Lei:Rinaldo:2015}, and \cite{Zhang:Zhou:2016}.
For a review of the subject, we refer to \cite{Bhattacharyya:Bickel:2016}.
However, how to validate the stochastic block model is a challenging problem and has not been addressed only until recently.
Specifically, \cite{Wang:Bickel:2017} developed a likelihood-based approach to test the model
and derived the asymptotic distribution of the log-likelihood ratio statistic under model misspecification
when the number of communities $k$ is fixed. \cite{Bickel:Sarkar:2015} used
the largest eigenvalue of the centered and
scaled adjacency matrix to test the Erd\H{o}s--R\'{e}nyi model and derived the asymptotic null distribution.
By extending their arguments,
\cite{Lei:2016} developed a goodness-of-fit test for stochastic block
models using the largest singular value of the centered and re-scaled adjacency matrix
and derived its asymptotic null distribution when the condition $k=o(n^{1/6})$ holds. It was also
acknowledged that it is difficult to extend these results to the more
flexible degree-corrected block model.
\cite{Karwa:Pati:Petrovic:Solus:Alexeev:Raic:Wilburne:Williams:Yan:2016} developed a finite-sample Monte Carlo goodness-of-fit test for the stochastic block model. The proposed test calculates goodness-of-fit statistics of graphs sampled from a conditional distribution given sufficient statistics of the stochastic block model; then the sample statistics are compared to the one calculated from the observed network, from which a naive $p$-value estimator is obtained. The proposed procedure is computationally expensive and there is no theoretical guarantee for the null distribution and asymptotic power of such finite-sample Monte Carlo tests.

In this article, we propose a novel goodness-of-fit test based on the maximum entry-wise deviation of the centered and re-scaled
adjacency matrix. We show that the asymptotic null distribution of the test statistic is a Gumbel distribution when $k=o(n/\log^2 n)$.
This condition implies that $k$ is allowed to grow linearly with
$n$ ignoring a logarithmic factor.  This kind of scenario has been referred by \cite{Rohe:Qin:Fan:2014} as the highest dimensional stochastic block model as the number of communities must be smaller than the number of nodes, and no reasonable model would allow $k$ to grow faster than that.
As a result, the proposed test significantly relaxes the condition that $k=o(n^{1/6})$ in \cite{Lei:2016}.
{Moreover, we show that the proposed test is asymptotically powerful against a class of alternatives. We also propose an augmented test statistic that improves the power of the goodness-of-fit test, while having the same asymptotic null distribution as the original test statistic.}
The maximum entry-wise deviation approach was first introduced by \cite{Jiang:2004} for testing the hypothesis $H_0: R=I$ vs $H_1: R \neq I$, where $R$ is a correlation matrix; therefore, the setting is quite different from ours.

The remainder of the article is organized as follows.  In Section \ref{section:test}, we introduce the new test statistic, and state its asymptotic null distribution and asymptotic power.
Further, we propose an augmented test statistic to improve the power of the test.
We extend our results to the degree-corrected stochastic block model in Section \ref{section:extension}.
Simulation studies and real-world data examples are given in Section \ref{section:simulation} and Section \ref{section:dataexample}, respectively. All proofs are collected in the supplementary materials.

\section{A new goodness-of-fit test for the stochastic block model}
\label{section:test}
Consider a stochastic block model on $n$ nodes with the membership vector $\sigma$ and probability matrix $B$.
For any fixed $(B,\sigma)$, the probability mass function for the adjacency matrix $A$ is
\[
P(A)=\prod_{1\leq i<j\leq n}B_{\sigma(i)\sigma(j)}^{A_{ij}}(1-B_{\sigma(i)\sigma(j)})^{(1-A_{ij})},
\]
and the corresponding log-likelihood under the stochastic block model can be written as
\[
\ell (A|B,\sigma)=\frac{1}{2}\sum_{u,v=1}^k(m_{uv}\log B_{uv}+(n_{uv}-m_{uv})\log(1-B_{uv})),
\]
where
$$
n_{uv}=\sum_{i=1}^n\sum_{j\neq i}\mathbf{1}\{\sigma(i)=u,\sigma(j)=v\}
~~\textrm{and}~~ m_{uv}=\sum_{i=1}^n\sum_{j\neq i}A_{ij}\mathbf{1}\{\sigma(i)=u,\sigma(j)=v\}.
$$
It is not difficult to see that given a number of communities $k_0$ and a membership vector $\sigma_0$, the maximum likelihood estimate of $B$ is given by
\begin{equation} \label{eq:bhat}
\widehat{B}^{\sigma_0}_{uv} = \left\{ \begin{array}{ll}
   \frac{ \sum_{i\in\sigma_0^{-1}(u), j\in\sigma_0^{-1}(v)} A_{ij} }{ \mid\sigma_0^{-1}(u)\mid \cdot \mid\sigma_0^{-1}(v)\mid }, & u\neq v, \\
   \frac{ \sum_{i\neq j\in\sigma_0^{-1}(u)} A_{ij} }{ \mid\sigma_0^{-1}(u)\mid \cdot (\mid\sigma_0^{-1}(u)\mid-1) }, & u=v,\\
   \end{array} \right.
\end{equation}
where $\sigma_0^{-1}(u)=\{i: 1\leq i\leq n, \sigma_0(i)=u\}$ and $\mid \sigma_0^{-1}(u)\mid$ is the number of nodes in block $u$.

Now given an observed adjacency matrix $A$, one may be interested in knowing whether $A$ can be well fitted by a stochastic block model with $k_0$ communities and/or a membership vector $\sigma_0$.
This leads to the following two hypothesis tests for fitness of the stochastic block model:
\begin{enumerate}
\item[(1)] {$H_0: k = k_0$ vs $H_1: k > k_0$}, and
\item[(2)] $H_0: \sigma=\sigma_0$ vs $H_1: \sigma\neq\sigma_0$,
\end{enumerate}
where we use $k$ and $\sigma$ to denote the true number of communities and the true membership vector respectively, and use $k_0$ and $\sigma_0$ to denote a hypothetical number of communities and a hypothetical membership vector respectively.
{Note in hypothesis test (1), we consider the one-sided alternative in which nodes are partitioned into less than $k$ communities (i.e., $k_0<k$).
For $k_0>k$, nodes are partitioned into more than $k$ communities.
In this case, goodness-of-fit tests may not have theoretical guarantee, as a stochastic block model with $k$ communities can also be reformulated as one with $k_0>k$ communities by artificially splitting one or more true communities.
As a result, we focus on the one-sided alternative $H_1: k > k_0$, similar to what has been considered in \cite{Lei:2016}, \cite{Chen:Lei:2017} and \cite{Wang:Bickel:2017}.
}

Let the centered and re-scaled adjacency matrix $\widetilde{A}$ be
\[
\widetilde{A}_{ij} = \left\{ \begin{array}{ll}
   \frac{ A_{ij}-\widehat{P}^{\sigma_0}_{ij} }{ \sqrt{(n-1)\widehat{P}^{\sigma_0}_{ij}(1-\widehat{P}^{\sigma_0}_{ij})} }, & i\neq j \\
   0, & i=j,
 \end{array} \right.
\]
where $\widehat{P}^{\sigma_0}_{ij}=\widehat{B}^{\sigma_0}_{\sigma_0(i)\sigma_0(j)}$, as defined in \eqref{eq:bhat}.
Under the null hypothesis $H_0: k=k_0$, $\sigma=\sigma_0$, if $k=o(n^{1/6})$, \cite{Lei:2016} showed that
\[
n^{2/3}(\lambda_1(\widetilde{A})-2)\stackrel{d}{\rightarrow} TW_1 \,\,\,\hbox{and}\,\,\,
n^{2/3}(-\lambda_n(\widetilde{A})-2)\stackrel{d}{\rightarrow} TW_1,
\]
where $TW_1$ denotes the Tracy-Widom distribution with index 1 and
$\lambda_i(A)$ denotes the $i$-th largest eigenvalue of the matrix $A$.
Further, to test (1), \cite{Lei:2016} proposed to obtain $\widehat{\sigma}$ using spectral clustering (under $k=k_0$)
and developed the following test statistic:
\[
T_{n,k_0}=\max[n^{2/3}(\lambda_1(\widetilde{A})-2), n^{2/3}(-\lambda_n(\widetilde{A})-2)],
\]
where $\sigma_0$ in $\widetilde{A}$ has been replaced by $\widehat{\sigma}$.
Note $T_{n,k_0}$ is a Bonferroni correction, and the corresponding level-$\alpha$ rejection rule is then
\[
\textrm{reject}\,\,\, H_0: k=k_0 \,\,\, \textrm{if} \,\,\, T_{n,k_0}\geq t_{1-\alpha/2},
\]
where $t_{\alpha}$ is the $\alpha$-th quantile of the $TW_1$ distribution for $\alpha\in (0,1)$.
As an improvement to many previous methods, the number of communities $k$ in \cite{Lei:2016} is allowed to grow as $n$ increases, but at the rate of $k=o(n^{1/6})$, which suggests that the test may not perform well when $k$ is large.

We aim to develop a new test statistic that allows $k$ to grow, up to a logarithm factor, linearly with $n$, and is able to test the goodness-of-fit of stochastic block models in both hypothesis tests (1) and (2).
Most existing work in the literature have only considered the hypothesis test (1), while as we will see, as a natural by-product of our result, we are also able to consider the hypothesis test (2), which is often of practical interest as well.
Moreover, the proposed test statistic can be extended to the degree-corrected stochastic block model.
Specifically, we propose a new test statistic based on the maximum entry-wise deviation:
\[
\begin{array}{lll}
L_n(k_0, \sigma_0) &\triangleq& \max_{1\leq i\leq n, 1\leq v\leq k_0} \mid\widehat{\rho}_{iv}\mid,
\end{array}
\]
where $\widehat{\rho}_{iv} = \frac{1}{ \sqrt{\mid\sigma_0^{-1}(v)/\{i\}\mid} } \sum_{j\in\sigma_0^{-1}(v)/\{i\}} \frac{ A_{ij}-\widehat{B}^{\sigma_0}_{\sigma_0(i)\sigma_0(j)} }{ \sqrt{\widehat{B}^{\sigma_0}_{\sigma_0(i)\sigma_0(j)}(1-\widehat{B}^{\sigma_0}_{\sigma_0(i)\sigma_0(j)})} }$, and $\sigma_0^{-1}(v)/\{i\}$ denotes the set of nodes, excluding node $i$, that belong to community $v$ in $\sigma_0$.

\subsection{The asymptotic null distribution}
\label{sec::null}

To derive the asymptotic distribution for $L_n(k_0, \sigma_0)$, we make the following assumptions:
\begin{enumerate}[({A}1)]
\item The entries of $B$ are uniformly bounded away from 0 and 1, and $B$ has no identical rows.
\item There exist $C_1>0$ and $C_2>0$ such that
\[
C_1n/k\leq \min_{1\leq u\leq k}\mid\sigma^{-1}(u)\mid\leq \max_{1\leq u\leq k}\mid\sigma^{-1}(u)\mid\leq C_2n^2/(k^2\log^2 n)
\]
for all $n$.
\end{enumerate}
\noindent Condition (A1) requires that the entries in the probability matrix $B$ are bounded away from 0 and 1, which was also considered in \cite{Lei:2016}. At the same time, Condition (A1) requires that $B$ is identifiable. Such a condition was considered in \cite{Wang:Bickel:2017} as well. Condition (A2) requires the size of the smallest community is at least proportional to $n/k$. This is a reasonable and mild condition; for example, it is satisfied almost surely if the membership vector $\sigma$ is generated from a multinomial distribution with $n$ trials and probability
$\pi=(\pi_1, \ldots, \pi_{k})$ such that $\min_{1\le u \le k} \pi_u \ge C_1/k$. Condition (A2) also places an upper bound on the largest community size. This is a reasonable condition as well and similar conditions have been considered by \cite{Zhang:Zhou:2016} and \cite{Gao:Ma:Zhang:Zhou:2018}.
The upper bound on the largest community size is used to control the maximum grouped bias between $\widehat{B}_{\sigma(i)\sigma(j)}$ and its population version $B_{\sigma(i)\sigma(j)}$, i.e.,
\[
\begin{array}{lll}
&\max_{1\leq i\leq n, 1\leq v\leq k}\mid\frac{1}{\sqrt{\mid\sigma^{-1}(v)/\{i\}\mid}}\sum_{j\in\sigma^{-1}(v)/\{i\}}\frac{B_{\sigma(i)\sigma(j)}-\widehat{B}_{\sigma(i)\sigma(j)}}{\sqrt{B_{\sigma(i)\sigma(j)}(1-B_{\sigma(i)\sigma(j)})}}\mid
\end{array}
\]
such that it converges in probability to $0$.

We now state the asymptotic properties of $L_n(k_0,\sigma_0)$ and delay the proof to the supplementary materials.

\begin{theorem}\label{theorem:null:a}
Suppose that conditions (A1) and (A2) hold.  Then under the null hypothesis $H_0: k=k_0$, $\sigma=\sigma_0$, as $n\rightarrow\infty$, if $k=o(n/\log^2 n)$,
we have
\begin{eqnarray} \label{eq:gumbel1}
\frac{L_n(k_0,\sigma_0)}{\sqrt{\log(2k_0n)}} &\stackrel{P}{\longrightarrow}& \sqrt{2}\,\,\,\,\, and \\
\lim_{n\to\infty}P(L_n^2(k_0,\sigma_0)-2\log(2k_0n)+\log\log(2k_0n)\leq y) &=&  \exp\{-\frac{1}{2\sqrt{\pi}}e^{-y/2}\}, \label{eq:gumbel2}
\end{eqnarray}
where the right hand side of \eqref{eq:gumbel2} is the cumulative distribution function of the Gumbel distribution with $\mu=-2\log(2\sqrt{\pi})$ and $\beta=2$.
\end{theorem}

Using the above theorem, we can carry out both hypothesis tests (1) and (2).
{To carry out hypothesis test (1), we need to first estimate the community membership $\widehat{\sigma}$ under $H_0: k=k_0$, and then compute
\[
T_n = L_n^2(k_0,\widehat{\sigma}) - 2\log(2k_0n) + \log\log(2k_0n).
\]
Assume that $\widehat{\sigma}$ is strongly consistent (i.e., $P(\widehat{\sigma}=\sigma_0)\rightarrow 1$). Following Theorem \ref{theorem:null:a}, we have that $T_n$ follows a Gumbel distribution with $\mu=-2\log(2\sqrt{\pi})$ and $\beta=2$. To carry out the test, we reject $H_0: k=k_0$, if {$T_n>t_{(1-\alpha)}$}, where $t_{\alpha}$ is the $\alpha$-th quantile of the Gumbel distribution with $\mu=-2\log(2\sqrt{\pi})$ and $\beta=2$.

In order to obtain the asymptotic null distribution of $T_n$ calculated with $\widehat{\sigma}$, the estimated $\widehat{\sigma}$ is required to be strongly consistent. This assumption is analogous to the strong consistency condition on $\widehat{\sigma}$ in \cite{Lei:2016} and the global optimum condition on the maximum likelihood estimation in \cite{Wang:Bickel:2017}. Under conditions (A1) and (A2), strong consistency (or exact recovery) is achievable when $k=o(n/\log^2 n)$ by Theorem 1.1 in \cite{Gao:Ma:Zhang:Zhou:2018}.
To achieve strong consistency, we consider the majority voting algorithm in \cite{Gao:Ma:Zhang:Zhou:2017}, initialized by spectral clustering \citep{Lei:Rinaldo:2015}.
Based on Theorem 4 in \cite{Gao:Ma:Zhang:Zhou:2017}, this procedure can achieve strong consistency when $k=o(n/\log^2 n)$.
Alternatively, to obtain $\widehat{\sigma}$, one may consider spectral clustering combined with the sample splitting method in \cite{Lei:Zhu:2017}, or the variational EM method in \cite{Daudin:Picard:Robin:2008}.
While the latter two methods perform well empirically, they do not have theoretic guarantee on strong consistency when $k$ diverges.
}

As for hypothesis test (2), since $\sigma_0$ gives a corresponding $k_0$, we can compute
\[
T_n = L_n^2(k_0, \sigma_0) - 2\log(2k_0n) + \log\log(2k_0n),
\]
and we reject $H_0: \sigma=\sigma_0$, if $T_n>t_{(1-\alpha)}$, where $t_{\alpha}$ is again the $\alpha$-th quantile of the Gumbel distribution with $\mu=-2\log(2\sqrt{\pi})$ and $\beta=2$.
In Section~\ref{section:simulation}, we carry out extensive simulation studies to investigate the finite sample performance of the two proposed tests of hypothesis.

\subsection{The asymptotic power}
\label{sec::power}
In this section, we study the asymptotic power of the proposed tests. To do so, we first define a class of alternatives. For a stochastic block model with true membership vector $\sigma$ and true probability matrix $B$, define probability matrix $B^{\sigma_0}$ with respect to a given membership vector $\sigma_0$ as
\[
B_{uv}^{\sigma_0} = \left\{ \begin{array}{ll}
   \frac{ \sum_{i\in\sigma_0^{-1}(u), j\in\sigma_0^{-1}(v)} B_{\sigma(i)\sigma(j)} }{ \mid\sigma_0^{-1}(u)\mid \cdot \mid\sigma_0^{-1}(v)\mid }, & u\neq v, \\
   \frac{ \sum_{i\neq j\in\sigma_0^{-1}(u)} B_{\sigma(i)\sigma(j)} }{ \mid\sigma_0^{-1}(u)\mid \cdot (\mid\sigma_0^{-1}(u)\mid-1) }, & u=v.\\
   \end{array} \right.
\]
From the above definition, we can see that $B^{\sigma}=B$. We introduce the following condition on $k_0$ and $\sigma_0$:
\begin{enumerate}[({A}2')]
\item There exist $C_1>0$ and $C_2>0$ such that
\[
C_1n/k_0\leq \min_{1\leq u\leq k_0}\mid\sigma_0^{-1}(u)\mid\leq \max_{1\leq u\leq k_0}\mid\sigma_0^{-1}(u)\mid\leq C_2n^2/(k_0^2\log^2 n),
\]
for all $n$.
\end{enumerate}
\noindent This condition is analogous to (A2) and as we have argued, is a reasonably mild condition on community sizes.

{Define the maximum grouped difference between $B$ and $B^{\sigma_0}$ as
\[
\begin{array}{lll}
\ell(k_0,\sigma_0)=\max_{1\leq i\leq n,1\leq v\leq k_0}\mid\frac{1}{\sqrt{\mid\sigma_0^{-1}(v)\mid}}\sum_{j\in\sigma_0^{-1}(v)}(B_{\sigma(i)\sigma(j)}-B^{\sigma_0}_{\sigma_0(i)\sigma_0(j)})\mid.
\end{array}
\]
Consider the following alternative class of number of communities and membership vectors:
\[
\begin{array}{lll}\label{eq:alternative}
&\mathcal{F}(k,\sigma,B)=\{(k_0,\sigma_0):k_0\le k,\, \ell(k_0,\sigma_0)/\sqrt{\log\, n}\longrightarrow\infty\}.
\end{array}
\]
The set $\mathcal{F}(k,\sigma,B)$ specifies that under the alternative, the maximum grouped difference between $B$ and $B^{\sigma_0}$ diverges faster than $\sqrt{\log n}$.
It can be seen that when $\sum_{j\in\sigma_0^{-1}(v)}(B_{\sigma(i)\sigma(j)}-B^{\sigma_0}_{\sigma_0(i)\sigma_0(j)})=O(|\sigma_0^{-1}(v)|)$ for some $i$ and $v$, under condition (A2') and $k_0=o(n/\log^2 n)$, we have that $\ell(k_0,\sigma_0)/\sqrt{\log\, n}\longrightarrow\infty$.
For example, when $k_0=k$, for an alternative $\sigma_0$ such that $B^{\sigma_0}\neq B$ (up to row/column permutations), we have $\sum_{j\in\sigma_0^{-1}(v)}(B_{\sigma(i)\sigma(j)}-B^{\sigma_0}_{\sigma_0(i)\sigma_0(j)})=O(|\sigma_0^{-1}(v)|)$ for some $i$ and $v$, and consequently $(k_0,\sigma_0)\in\mathcal{F}(k,\sigma,B)$.

Given $k$, $\sigma$, and $B$, it is straightforward to calculate $\ell(k_0,\sigma_0)$ for an alternative $(k_0,\sigma_0)$ and verify if it belongs to $\mathcal{F}(k,\sigma,B)$. We next provide some sufficient conditions for $(k_0,\sigma_0)\in\mathcal{F}(k,\sigma,B)$ when $k_0<k$.
\begin{corollary}\label{cor:alternative}
Suppose that conditions (A1) and (A2) hold. Consider the stochastic block model with $B$ and $\sigma$ from $multinomial\,(\pi_1, \ldots, \pi_{k})$. Let $B^{-}$ denote $B$ after removing the diagonal entries, i.e., $B^{-}_{u,\cdot}=(B_{uv})_{1\le v\le k, v\neq u}$, where $B^{-}_{u,\cdot}$ denotes the $u$th row of matrix $B^{-}$. For any $k_0<k$ and $\sigma_0$ satisfying $\sigma_0(i)=\sigma_0(j)$ if $\sigma(i)=\sigma(j)$, we have $(k_0,\sigma_0)\in\mathcal{F}(k,\sigma,B)$, if at least one of the following conditions holds for some $c_0>0$:
\begin{enumerate}[(i)]
\item $\min_{u\neq v}|B_{uu}-B_{vv}|>c_0$,
\item $\min_{u\neq v}\|B^{-}_{u,\cdot}-B^{-}_{v,\cdot}\|_{\infty}>c_0$, where $\|\cdot\|_{\infty}$ denotes the vector infinity norm,
\item $\min_{u\neq v}|\pi_{u}/\pi_{v}-1|>c_0$.
\end{enumerate}
\end{corollary}
\noindent The proof is collected in the supplementary materials.
In Corollary \ref{cor:alternative}, we focus on the merged alternatives (i.e., communities in $\sigma$ are merged to form communities in $\sigma_0$) to reduce the number of possible alternatives in developing the theoretical result, similar to that in \cite{Wang:Bickel:2017}.
Condition (i) specifies that the absolute differences between diagonal entries in $B$ are lower bounded, condition (ii) specifies that the differences, in terms of the infinity norm, between rows in $B^{-}$ are lower bounded, and condition (iii) specifies that the differences between elements in $(\pi_1, \ldots, \pi_{k})$ are lower bounded.
These conditions cover a large class of stochastic block models.
Next, we discuss the asymptotic power of our proposed test against alternatives in $\mathcal{F}(k,\sigma,B)$.
The following theorem provides a lower bound on the growth rate of the test statistic under an alternative $(k_0,\sigma_0)\in\mathcal{F}(k,\sigma,B)$.
\begin{theorem}\label{theorem:alternative:a}
Suppose that conditions (A1) and (A2') hold. For any alternative $(k_0,\sigma_0)\in\mathcal{F}(k,\sigma,B)$, let $T_n=L_n^2(k_0,\sigma_0)-2\log(2k_0n)+\log\log(2k_0n)$. If $k_0=o(n/\log^2 n)$, we have
\begin{eqnarray} \label{eq:gumbel3}
P(T_n\ge c_1\log(n))\rightarrow 1,
\end{eqnarray}
for some positive constant $c_1$.
\end{theorem}
\noindent The proof is collected in the supplementary materials. Theorem \ref{theorem:alternative:a} shows that the growth rate of $T_n$ under the alternative is at least $\log(n)$. The asymptotic null distribution in Theorem \ref{theorem:null:a} and the growth rate under the alternative suggest that the null and the alternative hypotheses are well separated, and our proposed test is asymptotically powerful against $(k_0,\sigma_0)\in\mathcal{F}(k,\sigma,B)$. Specifically, our proposed test is asymptotically powerful when $k_0<k$, if at least one of the conditions in Corollary \ref{cor:alternative} holds.

Notably, however, under the planted partition model (i.e., $B_{uu}=p$ and $B_{uv}=q, u\neq v$ for some $0\le q<p \le 1$) with equal sized communities, some straightforward algebra shows that $\ell(k_0,\sigma_0)=0$ for any $(k_0,\sigma_0)$ satisfying $k_0<k$ and $\sigma_0(i)=\sigma_0(j)$ if $\sigma(i)=\sigma(j)$.
Consequently, such alternatives do not belong to $\mathcal{F}(k,\sigma,B)$.
Additionally, one can verify that our test is not powerful against such alternatives under the planted partition model with equal sized communities.
For example, consider a simple case with $k=2$ and $\pi_1=\pi_2$. Under $k_0=1$, we have $B^{\sigma_0}=(p+q)/2$, and the entry-wise deviation is calculated as
\begin{eqnarray*}
&\rho_{i1}=\frac{1}{\sqrt{n}}\left|\sum_{j\in\sigma_0^{-1}(1)}\frac{A_{ij}-\frac{p+q}{2}}{\sqrt{\frac{p+q}{2}(1-\frac{p+q}{2})}}\right|=\frac{1}{\sqrt{n}}\frac{\left|\sum_{j\in\sigma^{-1}(1)}(A_{ij}-p)+\sum_{j\in\sigma^{-1}(2)}(A_{ij}-q)\right|}{\sqrt{\frac{p+q}{2}(1-\frac{p+q}{2})}}.
\end{eqnarray*}
It can be seen that this entry-wise deviation is not well separated from those calculated under the null, i.e., $\frac{1}{\sqrt{n/2}}\frac{\left|\sum_{j\in\sigma^{-1}(1)}(A_{ij}-p)\right|}{\sqrt{p(1-p)}}$ and $\frac{1}{\sqrt{n/2}}\frac{\left|\sum_{j\in\sigma^{-1}(2)}(A_{ij}-q)\right|}{\sqrt{q(1-q)}}$, and our proposed test is not powerful. The above calculation can be generalized to the cases where $k\ge2$ and $k_0<k$.
Returning to the case of $k=2$ and $k_0=1$, it is straightforward to show that when $|\pi_1/\pi_2-1|>c_0$ for some $c_0>0$, the growth rate of $\max_{i,v}\rho_{i1}$ is $\sqrt{n}$.
Consequently, the null and the alternative are well separated, and our proposed test is powerful.
More generally, for $k\ge2$ and $k_0<k$, if $\min_{u\neq v}|\pi_{u}/\pi_{v}-1|>c_0$ for some $c_0>0$, our proposed test is powerful (see Corollary \ref{cor:alternative}).

\subsection{An augmented test statistic}\label{sec:augment}
In this section, we discuss a practical solution to improve the power of the proposed test for hypothesis test (1) under the planted partition model with equal sized communities.
Consider a planted partition model with $n$ nodes, $k$ equal sized communities, and within and between community connecting probabilities $p$ and $q$, respectively.
We propose adding a community of size $n/(2k)$ to the model.
For the added community, we let the within and between community connecting probabilities be $p'$ and $q'$, respectively. Note that $(p', q')$ can be the same as or different from $(p,q)$.
For this new model with $k+1$ communities, by Theorem \ref{theorem:null:a}, the asymptotic null distribution of the test statistic still follows a Gumbel distribution.
Moreover, under an alternative $k_0<k+1$, if the added small community is merged with others to form a new community in $\sigma_0$, we have $\ell(k_0,\sigma_0)\in\mathcal{F}(k,\sigma,B)$ and our proposed test is powerful.
For spectral clustering based algorithms, such as that in \cite{Gao:Ma:Zhang:Zhou:2017}, it is reasonable to assume that the added small community is merged with others in $\sigma_0$ when $k_0<k+1$, as this tends to lead to smaller within-cluster sum of squares.

The discussion above implies that when carrying out hypothesis test (1), an additional community can be added to the observed network to improve the power of our proposed test.
We refer to the test statistic calculated with the added community as the \textit{augmented test statistic}. Denote $k_0^+=k_0+1$.
For a given adjacency matrix $A$ and null hypothesis $H_0:\,k=k_0,\,\sigma=\sigma_0$, the augmented test statistic is calculated through the following steps:
\begin{enumerate}[(1)]
\item Calculate $\widehat B$ using \eqref{eq:bhat}.
\item Add a $k_0^+$th community of size $n_{k_0^+}=\min_{1\le u\le k_0} |\sigma_0^{-1}(u)|/2$ to the observed network.
For the added community, let the within and between community connecting probabilities be $\max_{1\le u\le k_0}\widehat B_{uu}$ and $\min_{u\neq v}\widehat B_{uv}/2$, respectively.
\item Calculate the size $n^+$ and the adjacency matrix $A^+$ of the network from step (2). With the membership vector $\sigma^+_0=(\sigma_0,\underbrace{k_0^+,\ldots,k_0^+}_{n_{k_0^+}})$, calculate the probability matrix $\widehat B^+$.
\item The augmented test statistic is calculated as
$$
T^+_n= L_n^2(k_0^+, \sigma^+_0) - 2\log(2k_0^+n^+) + \log\log(2k_0^+n^+).
$$
\end{enumerate}
To carry out hypothesis test (1), we reject the null hypothesis if $T^+_n>t_{(1-\alpha)}$, where $t_{\alpha}$ is the $\alpha$-th quantile of the Gumbel distribution with $\mu=-2\log(2\sqrt{\pi})$ and $\beta=2$.

Note that the asymptotic null distribution of the augmented test statistic $T^+_n$ is the same as $T_n$, provided that the added community satisfies Conditions (A1) and (A2). Under Conditions (A1) and (A2), other procedures (e.g., different size or connecting probabilities) for adding the additional community are also feasible.
We adopt the above procedure as it is easy to implement and shows good empirical performance in our numerical studies.

}

\section{Extension to the degree-corrected stochastic block model}
\label{section:extension}
It has been observed that a typical real-world network often contains a few high-degree ``hub'' nodes which have many edges
and many low-degree nodes that have few edges. The stochastic block model, however, does not accommodate such heterogeneity.
To incorporate the degree heterogeneity of nodes for community detection, \cite{Karrer:Newman:2011} proposed the degree-corrected stochastic block model. Specifically,
the degree-corrected stochastic block model assumes that $P(A_{ij}=1\mid \sigma(i)=u,\sigma(j)=v)=\omega_i\omega_jB_{uv}$, where
$\mathbf{\omega}=(\omega_i)_{1\leq i\leq n}$ are a set of node degree parameters measuring
the degree variation.
For identifiability of the model,
we use the following constraint for the degree-corrected stochastic block model:
\begin{enumerate}[({A}3)]
\item $\sum_i\omega_i\mathbf{1}\{\sigma(i)=u\}=|\sigma^{-1}(u)|$ for each community $1\leq u\leq k$.
\end{enumerate}

To develop a goodness-of-fit test for the degree-corrected stochastic block model, we consider two cases: (1) $\omega$ is known, and (2) $\omega$ is unknown. We first consider the case where $\omega$ is known.
In this case, we propose the following test statistic:
\[
\begin{array}{lll}
L_{n1}(k_0, \sigma_0) &\triangleq& \max_{1\leq i\leq n,1\leq v\leq k_0} \mid\widehat{\tau}_{iv}\mid,
\end{array}
\]
where $\widehat{\tau}_{iv} = \frac{1}{ \sqrt{\mid\sigma_0^{-1}(v)/\{i\}\mid} } \sum_{j\in\sigma_0^{-1}(v)/\{i\}} \frac{ A_{ij}-\omega_i\omega_j\widehat{B}^{\sigma_0}_{\sigma_0(i)\sigma_0(j)} }{ \sqrt{\omega_i\omega_j\widehat{B}^{\sigma_0}_{\sigma_0(i)\sigma_0(j)}(1-\omega_i\omega_j\widehat{B}^{\sigma_0}_{\sigma_0(i)\sigma_0(j)})} }$.\\

To derive the asymptotic distribution of $L_{n1}(k_0,\sigma_0)$, we make the following additional assumption:
\begin{enumerate}[({A}4)]
\item The entries of $(\omega_i\omega_jB_{\sigma(i)\sigma(j)})_{1\leq i\leq n, 1\leq j\leq n}$ are uniformly bounded away from 0 and 1.
\end{enumerate}
We now state the asymptotic properties of $L_{n1}(k_0,\sigma_0)$.
\begin{theorem}\label{theorem:null:b}
Suppose that conditions (A2)-(A4) hold.  Then under the null hypothesis $H_0: k=k_0$, $\sigma=\sigma_0$, as $n\rightarrow\infty$, if $k=o(n/\log^2 n)$, we have
\begin{eqnarray*}
\frac{L_{n1}(k_0,\sigma_0)}{\sqrt{\log(2k_0n)}} &\stackrel{P}{\longrightarrow}& \sqrt{2}\,\,\,\,\, and \\
\lim_{n\to\infty}P(L_{n1}^2(k_0,\sigma_0) - 2\log(2k_0n) + \log\log(2k_0n)\leq y) &=&  \exp\{-\frac{1}{2\sqrt{\pi}}e^{-y/2}\}.
\end{eqnarray*}
\end{theorem}

Note that $E(\frac{A_{ij}-\omega_i\omega_jP_{ij}}{\sqrt{\omega_i\omega_jP_{ij}(1-\omega_i\omega_jP_{ij})}})=0$ and $E(\frac{A_{ij}-\omega_i\omega_jP_{ij}}{\sqrt{\omega_i\omega_jP_{ij}(1-\omega_i\omega_jP_{ij})}})^2=1$, which is analogous to the result under the stochastic block model, in which $E(\frac{A_{ij}-B_{\sigma(i)\sigma(j)}}{\sqrt{B_{\sigma(i)\sigma(j)}(1-B_{\sigma(i)\sigma(j)})}})=0$ and $E(\frac{A_{ij}-B_{\sigma(i)\sigma(j)}}{\sqrt{B_{\sigma(i)\sigma(j)}(1-B_{\sigma(i)\sigma(j)})}})^2=1$.
Henceforth, the proof of Theorem~\ref{theorem:null:b} is very similar to that of Theorem \ref{theorem:null:a}, and we omit the details in the article.
Using the result in the above theorem, we can carry out hypothesis tests (1) and (2) using the test $\Phi_{\alpha}$ defined as
\[
\Phi_{\alpha}=I(T_{n1}>t_{(1-\alpha)}),
\]
where $T_{n1} = L_{n1}^2(k_0, \sigma_0) - 2\log(2k_0n) + \log\log(2k_0n)$ and $t_{\alpha}$ is the $\alpha$-th quantile of the Gumbel distribution with $\mu=-2\log(2\sqrt{\pi})$ and $\beta=2$.
{To estimate $\widehat\sigma$, we adopt the regularized spherical spectral clustering algorithm in \cite{Lei:Rinaldo:2015}. Other methods such as the SCORE algorithm in \cite{{Jin:2015}} and the normalized neighbor voting procedure in \cite{Gao:Ma:Zhang:Zhou:2018} can also be considered. }
Following similar arguments as in the case of stochastic block model, it can also be shown that the test $\Phi_{\alpha}$ is powerful against a class of alternatives defined similarly as in \eqref{eq:alternative}.
{Following similar arguments as in Section \ref{sec::power}, we can show that when the communities are equal sized and $B_{uu}=p$ and $B_{uv}=q, u\neq v$ for some $0\le q<p \le 1$, our proposed test is not powerful when $k_0<k$. To overcome this challenge, we propose an augmented test statistic for the degree-corrected stochastic block model. The calculation of this augmented test statistic is similar to that under the stochastic block model, and we include the computational details in the supplementary materials.
}

If $\omega$ is unknown, we can plug in its estimate for $L_{n1}(k_0,\sigma_0)$. Similar to \cite{Karrer:Newman:2011}, we replace the Bernoulli distribution of $A_{ij}$ by the Poisson distribution with the mean $\omega_i \omega_j B_{uv}$.
As discussed in \cite{Zhao:Levina:Zhu:2012}, there is no practical difference in performance between the log-likelihood and its slightly more elaborate version
based on the Bernoulli observations.
The reason is that the Bernoulli distribution with a small mean can be well approximated by a Poisson distribution. One advantage of using the Poisson distribution is that it greatly simplifies the calculation.
Another advantage is that it admits networks containing both multi-edges and self-edges. Specifically, for any fixed $(B,\omega, \sigma)$, the log-likelihood of observing the adjacency matrix $A$ under the degree-corrected stochastic block model can be written as
\[
\ell (A|B,\omega, \sigma)=\sum_{1\leq i\leq n}d_i\log\omega_i+\frac{1}{2}\sum_{u, v=1}^k(m_{uv}\log B_{uv}-n_{uv}B_{uv}),
\]
where $d_i=\sum_{1\leq j\leq n}A_{ij}$, and $m_{uv}$ and $n_{uv}$ are defined the same as before.
It is not difficult to show that given $\sigma_0$, the maximum likelihood estimate of the parameter $\omega$ is given by
$\widehat{\omega}_i=|\sigma_0^{-1}(u)|d_i/\sum_{j:\sigma_0(j)=\sigma_0(i)}d_j$.
Then the proposed plug-in test statistic is given by
\begin{equation}\label{eq:dcsbm}
\begin{array}{lll}
L_{n2}(k_0,\sigma_0) &\triangleq& \max_{1\leq i\leq n,1\leq v\leq k_0} \mid\widehat{\tau}_{iv}\mid,
\end{array}
\end{equation}
where $\widehat{\tau}_{iv} = \frac{1}{\sqrt{\mid\sigma_0^{-1}(v)/\{i\}\mid}} \sum_{j\in\sigma_0^{-1}(v)/\{i\}}  \frac{ A_{ij}-\widehat{\omega}_i\widehat{\omega}_j
\widehat{B}^{\sigma_0}_{\sigma_0(i)\sigma_0(j)} }{ \sqrt{\widehat{\omega}_i\widehat{\omega}_j\widehat{B}^{\sigma_0}_{\sigma_0(i)\sigma_0(j)}(1-\widehat{\omega}_i\widehat{\omega}_j\widehat{B}^{\sigma_0}_{\sigma_0(i)\sigma_0(j)})} }$.\\

When $\omega$ is unknown, it is very challenging to derive the asymptotic distribution of $L_{n2}(k_0,\sigma_0)$, due to the complex dependency between the centered and re-scaled entries in $\widehat{\tau}_{iv}$. {We perform simulation studies and find that the empirical distribution of $L_{n2}^2(k_0,\sigma_0)-2\log(2k_0n)+\log\log(2k_0n)$ deviates from the Gumbel distribution by a location and scale shift (see Figure \ref{figure-b}). This shift is especially large when the number of communities $k$ is small. As a practical solution, in Section \ref{sec:bootdcsbm}, we describe a bootstrap correction procedure. With the bootstrap corrected test statistic, both hypothesis tests (1) and (2) can be carried out, similar to what have been done in Section \ref{section:test}.}

\renewcommand{\thefootnote}{\arabic{footnote}}
\section{Simulation studies}
\label{section:simulation}

In this section, we carry out extensive simulation studies to evaluate the performance of the proposed test statistic.
We consider both the stochastic block model and the degree-corrected stochastic block model. {In the stochastic block model setting, the majority voting algorithm in \cite{Gao:Ma:Zhang:Zhou:2017}, initialized by spectral clustering is used to obtain the community membership, whereas in the degree-corrected stochastic block model setting, the regularized spherical spectral clustering algorithm in \cite{Lei:Rinaldo:2015} is employed.
For the stochastic block model, we consider the test statistic $T_{n}=L_{n}^2(k_0,\sigma_0)-2\log(2k_0n)+\log\log(2k_0n)$, and the augmented test statistic $T^+_{n}$ proposed in Section \ref{sec:augment}.
For the degree-corrected stochastic block model, we consider $T_{n2}=L_{n2}^2(k_0,\sigma_0)-2\log(2k_0n)+\log\log(2k_0n)$, and the augmented test statistic $T^+_{n2}$ proposed in Section \ref{sec:augmentdcsbm}.}
In our comparative simulation studies, \cite{Lei:2016}, \cite{Karwa:Pati:Petrovic:Solus:Alexeev:Raic:Wilburne:Williams:Yan:2016} and our method are all implemented in R\footnote{We obtained the code for \cite{Lei:2016} from the author's website and implemented the code for \cite{Karwa:Pati:Petrovic:Solus:Alexeev:Raic:Wilburne:Williams:Yan:2016} by ourselves following the algorithm proposed in the article.}.

\medskip
\noindent \textbf{Simulation 1. The null distribution under the stochastic block model and a bootstrap correction.}

\noindent {In this simulation, we examine the finite sample null distribution of the test statistic $T_n$ and verify the result in Theorem \ref{theorem:null:a}.
As the speed of convergence to a Gumbel distribution may be slow, one may consider a finite sample bootstrap correction. Such a finite-sample correction was first proposed in \cite{Bickel:Sarkar:2015} and later considered in \cite{Lei:2016}. Here, we extend their ideas to our setting.

For an adjacency matrix $A$ and null hypothesis $k=k_0,\sigma=\sigma_0$, the bootstrap corrected goodness-of-fit test statistic is calculated as the following:
\begin{enumerate}
\item Calculate $\widehat B$ using \eqref{eq:bhat}. Calculate $T_n$ using $A$ and $(\widehat B,\sigma_0)$.
\item For $m=1,\ldots,M$, generate $A^{(m)}$ from the stochastic block model $(\widehat B,\sigma_0)$, and calculate $T_n^{(m)}$ using $A^{(m)}$ and $(\widehat B,\sigma_0)$.
\item Using $(T_n^{(m)}:1\le m\le M)$, estimate the location and scale parameters  $\widehat\mu$ and $\widehat\beta$ of the Gumbel distribution using maximum likelihood estimation.
\item The bootstrap corrected test statistic is calculated as
$$
T_{n,\text{boot}}=\mu+\beta\left(\frac{T_n-\widehat\mu}{\widehat \beta}\right),
$$
where $\mu=-2\log(2\sqrt{\pi})$ and $\beta=2$.
\end{enumerate}
Since the limiting distribution of the test statistic is provably Gumbel, finite sample corrections can be made inexpensively by generating a small number of bootstrap samples to estimate the location and scale parameters. In all of our simulations, we use $M=100$.

\begin{figure}[!t]
\centering
\includegraphics[trim=0 15mm 0 0, scale=0.6]{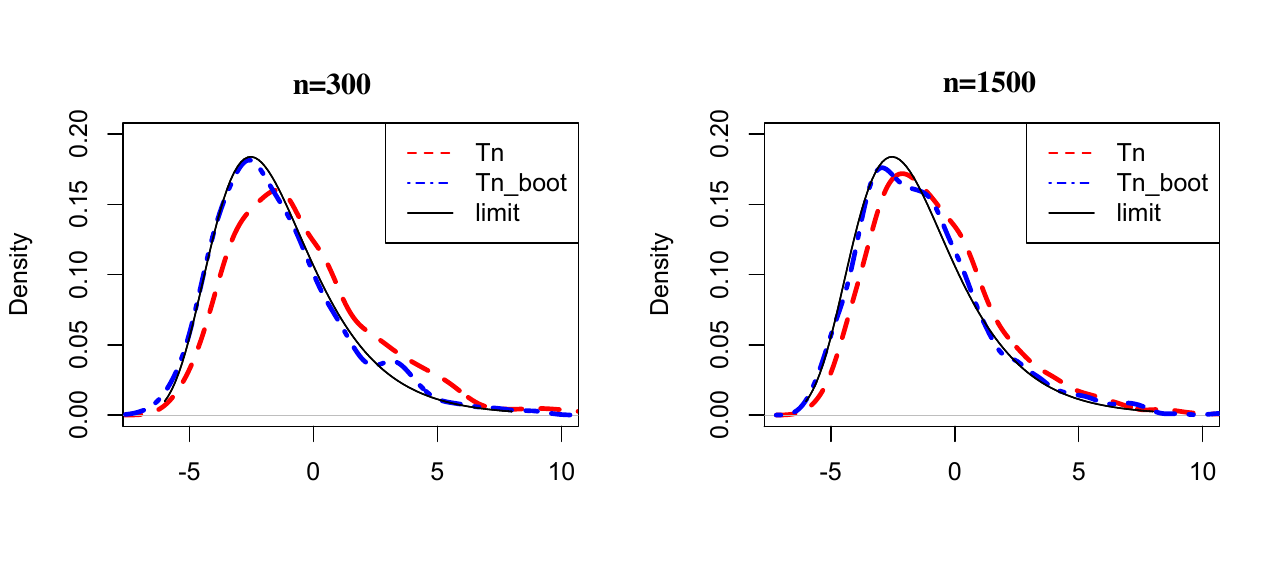}
\caption{Null densities under the stochastic block model in Simulation 1 with $n=300$ (left plot) and $n=1500$ (right plot). The red dashed lines, blue dash-dotted lines and black solid lines show the densities of the test statistic $T_n$, the bootstrap corrected test statistic $T_{n,\text{boot}}$ and the theoretical limit, respectively. }
\label{figure-a}
\end{figure}

In Figure \ref{figure-a}, we plot the distribution of ${T_n}$ with and without bootstrap corrections from 1000 data replications.
In this simulation, we set $k=k_0=3$ with $\pi_1=\pi_2=\pi_2=1/3$. The edge probability between communities $u$ and $v$ is $B_{uv}=0.1(1+2\times\mathbf{1}(u=v))$.
We consider sample sizes $n=300$ and $n=1500$.
It can be seen that the finite sample null distribution of $T_n$ deviates slightly from the limiting distribution when $n = 300$, and the difference is much reduced when $n = 1500$. When bootstrap correction is considered, the finite sample null distribution is close to the limit even when $n = 300$.}

\medskip
\noindent \textbf{Simulation 2. Hypothesis test (1) under the stochastic block model.}

In the stochastic block model setting, we consider the hypothesis test
$$H_0: k=k_0 \,\,\,vs \,\,\,H_1: k>k_0.$$
{We first compare the performance of the test statistic $T_n$, the augmented test statistic $T_n^+$ and the bootstrap corrected augmented test statistic $T^+_{n, \text{boot}}$ under varying $k$ and $k_0$. We set the edge probability between communities $u$ and $v$ as $0.1(1+4\times\mathbf{1}(u=v))$, and let the size of each block be 200. Table \ref{tab:compare} reports the result from 100 data replications. While the type I errors from all three test statistics are close to the nominal level, $T^+_{n, \text{boot}}$'s type I errors are closer to the nominal level when $k$ is large.
As this simulation setting considers a planted partition model with equal sized communities, $T_n$ does not have good power. This agrees with our theoretical results in Section \ref{sec::power}.
It is seen that, with the augmentation, the power from both $T^+_{n}$ and $T^+_{n, \text{boot}}$ improve significantly.
\begin{table}[!t]
\caption{Proportion of rejection at nominal level $\alpha=0.05$ for hypothesis test $H_0: k=k_0$ vs $H_1: k>k_0$. Each community has 200 nodes and $B_{uv}=0.1(1+4\times\mathbf{1}(u=v))$.}
\vspace{0.15in}
\centering
\scalebox{0.7}{
\renewcommand\arraystretch{1.5}
\begin{tabular}{c|ccccc|ccccc|ccccc}
\hline
& \multicolumn{5}{c|}{$T_{n}$} & \multicolumn{5}{c|}{$T^+_{n}$} & \multicolumn{5}{c}{$T^+_{n, \text{boot}}$}  \\
\hline
$k$                            &2 & 4 & 6 & 8 & 10                                               &2 & 4 & 6 & 8 & 10                                  &2 & 4 & 6 & 8 & 10   \\
\hline
$k_0=2$                    &\textbf{0.03}&0.09&0.43&0.66&0.82&                  \textbf{0.06}&1.00&1.00&1.00&1.00&                   \textbf{0.05}&1.00&1.00&1.00&1.00\\
$k_0=4$                    &*&\textbf{0.04}&0.08&0.42&0.88&                         *&\textbf{0.08}&1.00&1.00&1.00&                         *&\textbf{0.05}&1.00&1.00&1.00\\
$k_0=6$                    &*&*&\textbf{0.08}&0.14&0.28&                                 *&*&\textbf{0.08}&1.00&1.00&                                 *&*&\textbf{0.04}&1.00&1.00\\
$k_0=8$                    &*&*&*&\textbf{0.08}&0.14&                                         *&*&*&\textbf{0.09}&1.00&                                       *&*&*&\textbf{0.05}&1.00\\
$k_0=10$                  &*&*&*&*&\textbf{0.08}&                                                 *&*&*&*&\textbf{0.10}&                                             *&*&*&*&\textbf{0.04}\\
\hline
\end{tabular}}\label{tab:compare}
\end{table}

Next, we compare our method with \cite{Lei:2016}. For their test statistic, we also use the bootstrap correction procedure suggested in their work, and this test statistic is referred to as $\text{Lei}_{\text{boot}}$.
We fix the network size at $n=3000$ and let both $k$ and $k_0$ vary.
Table \ref{tab:b} reports the results from $T^+_{n, \text{boot}}$ and $\text{Lei}_{\text{boot}}$ from 200 data replications.
It can be seen from Table \ref{tab:b} that the two tests have comparable type I errors when $k$ is small (i.e., $k\le 5$).
However, when $k$ is large (i.e., $k\ge 10$), the type I errors from $T^+_{n, \text{boot}}$ are much closer to the nominal level.
This agrees with our theoretical finding that our proposed test allows $k$ to grow at a much faster rate than that of \cite{Lei:2016}.
Moreover, it is seen that both tests have good power. Specifically, both tests are powerful against the Erd\H{o}s--R\'{e}nyi model alternative (i.e., $k_0=1$) when $k\ge 2$.

\begin{table}[!t]
\caption{Proportion of rejection at nominal level $\alpha=0.05$ for $H_0: k=k_0$ vs $H_1: k>k_0$. The network size is $n=3000$ with equal sized communities, and $B_{uv}=0.1(1+4\times\mathbf{1}(u=v))$.}
\vspace{0.15in}
\centering
\scalebox{0.75}{
\renewcommand\arraystretch{1.5}
\begin{tabular}{c|ccccccc|ccccccc}
\hline
& \multicolumn{7}{c|}{$T^+_{n,\text{boot}}$} & \multicolumn{7}{c}{$\text{Lei}_{\text{boot}}$} \\
\hline
$k$                            &1 & 2 & 3 & 5 & 10 & 20 & 30                 &1& 2 & 3 & 5 & 10 & 20 & 30  \\
\hline
$k_0=1$                    &\textbf{0.03}&1.00&1.00&1.00&1.00&1.00&1.00&      \textbf{0.04}&1.00&1.00&1.00&1.00&1.00&1.00\\
$k_0=2$                    &*&\textbf{0.06}&1.00&1.00&1.00&1.00&1.00&            *&\textbf{0.04}&1.00&1.00&1.00&1.00&1.00\\
$k_0=3$                    &*&*&\textbf{0.06}&1.00&1.00&1.00&1.00&                  *&*&\textbf{0.02}&1.00&1.00&1.00&1.00\\
$k_0=5$                    &*&*&*&\textbf{0.05}&1.00&1.00&1.00&                        *&*&*&\textbf{0.03}&1.00&1.00&1.00\\
$k_0=10$                  &*&*&*&*&\textbf{0.06}&1.00&1.00&                              *&*&*&*&\textbf{0.46}&1.00&1.00\\
$k_0=20$                  &*&*&*&*&*&\textbf{0.04}&1.00&                                   *&*&*&*&*&\textbf{0.82}&1.00\\
$k_0=30$                  &*&*&*&*&*&*&\textbf{0.04}&                                         *&*&*&*&*&*&\textbf{0.98}\\
\hline
\end{tabular}}\label{tab:b}
\end{table}

We have also run comparative simulations with sparser networks, networks with unbalanced community sizes and networks with randomly generated $B$. In the interest of space, we report these additional results in the supplementary materials.
}

\medskip
\noindent \textbf{Simulation 3. Hypothesis test (2) under the stochastic block model.}

\noindent In the stochastic block model setting, we also consider the hypothesis test
$$H_0: \sigma=\sigma_0 \,\,\,vs \,\,\,H_1: \sigma\neq\sigma_0.$$
We use the true number of communities $k$ when we obtain the membership vector $\sigma_0$.
We investigate the probability of type I error of the test statistic $T_{n}$ and $T_{n,\text{boot}}$.
The network size $n$ is the same as in Simulation 2.
The edge probability between communities $u$ and $v$ is $0.1(1+2\times\mathbf{1}(u=v))$.
Each simulation is repeated 200 times.
The simulation results are given in Table \ref{tab:c}.
It can be seen from this table that the probabilities of type I error of both $T_{n}$ and $T_{n,\text{boot}}$ are close to the nominal level, with $T_{n,\text{boot}}$ having a slightly smaller type I error when $k$ is large.
We also compare our method with \cite{Karwa:Pati:Petrovic:Solus:Alexeev:Raic:Wilburne:Williams:Yan:2016}.
We can see that the estimated type I errors from our tests are much closer to the nominal level than that of \cite{Karwa:Pati:Petrovic:Solus:Alexeev:Raic:Wilburne:Williams:Yan:2016}.
\begin{table}[!t]
\caption{Proportion of rejection at nominal level $\alpha=0.05$ for hypothesis test $H_0: \sigma=\sigma_0$ vs $H_1: \sigma\neq\sigma_0$ under settings in Simulation 3.
\vspace{0.15in}
\label{tab:c}}
\centering
\scalebox{1}{
\renewcommand\arraystretch{1.2}
\begin{tabular}{c|ccccccc}
\hline
$k=k_0$ & 2 & 3 & 4 & 5 & 6 & 7 & 8 \\
\hline
$T_n$ & 0.05 & 0.05 & 0.07 & 0.07 & 0.09 & 0.07 & 0.10 \\
$T_{n,\text{boot}}$ & 0.04 & 0.08 & 0.05 & 0.03 & 0.06 & 0.03 & 0.04 \\
\cite{Karwa:Pati:Petrovic:Solus:Alexeev:Raic:Wilburne:Williams:Yan:2016} & 0.19 & 0.10 & 0.14  & 0.15 & 0.15  & 0.15 & 0.20 \\ \hline
\end{tabular}
}
\end{table}

{We have also run simulations to examine the power of the test when a proportion of the labels in $\sigma$ are corrupted. We find that our test is powerful under this setting. In the interest of space, we report these additional results in the supplementary materials.}

\medskip
\noindent \textbf{Simulation 4. The null distribution under the degree-corrected stochastic block model and a bootstrap correction.}
\noindent {
In this simulation, we examine the finite sample null distribution of the test statistic $T_{n2}$ under the degree-corrected stochastic block model. Similar to Simulation 1, we also consider a finite sample bootstrap correction. The calculation for the bootstrap corrected test statistic $T_{n2,\text{boot}}$ is similar to that in Simulation 1 and we include the computational details in the supplementary materials.}

To generate the degree parameters $\omega$, we follow the approach in \cite{Zhao:Levina:Zhu:2012}. The identifiability constraint  $\sum_i\omega_i\mathbf{1}\{\sigma(i)=u\}=|\sigma^{-1}(u)|$ for each community $1\leq u\leq k$ is replaced by the requirement that the $\omega_i$ be independently generated from a distribution with unit expectation, i.e.
$$\omega_i=\left\{
        \begin{array}{ll}
          \eta_i, & \hbox{w.p. 0.8}, \\
          9/11, & \hbox{w.p. 0.1}, \\
          13/11, & \hbox{w.p. 0.1},
        \end{array}
      \right.
$$
where $\eta_i$ is uniformly distributed on the interval $[\frac{4}{5},\frac{6}{5}]$.
{In this simulation, we consider $k=k_0=3$ with $\pi_u=1/3$, $u=1,\ldots,3$, and $k=k_0=5$ with $\pi_u=1/5$, $u=1,\ldots,5$.
The edge probability between communities $u$ and $v$ is $B_{uv}=0.1(1+2\times\mathbf{1}(u=v))$.
We consider sample sizes $n=300, 500$ and $1500$.

\begin{figure}[!t]
\centering
\includegraphics[scale=0.5, trim=0 10mm 0 0]{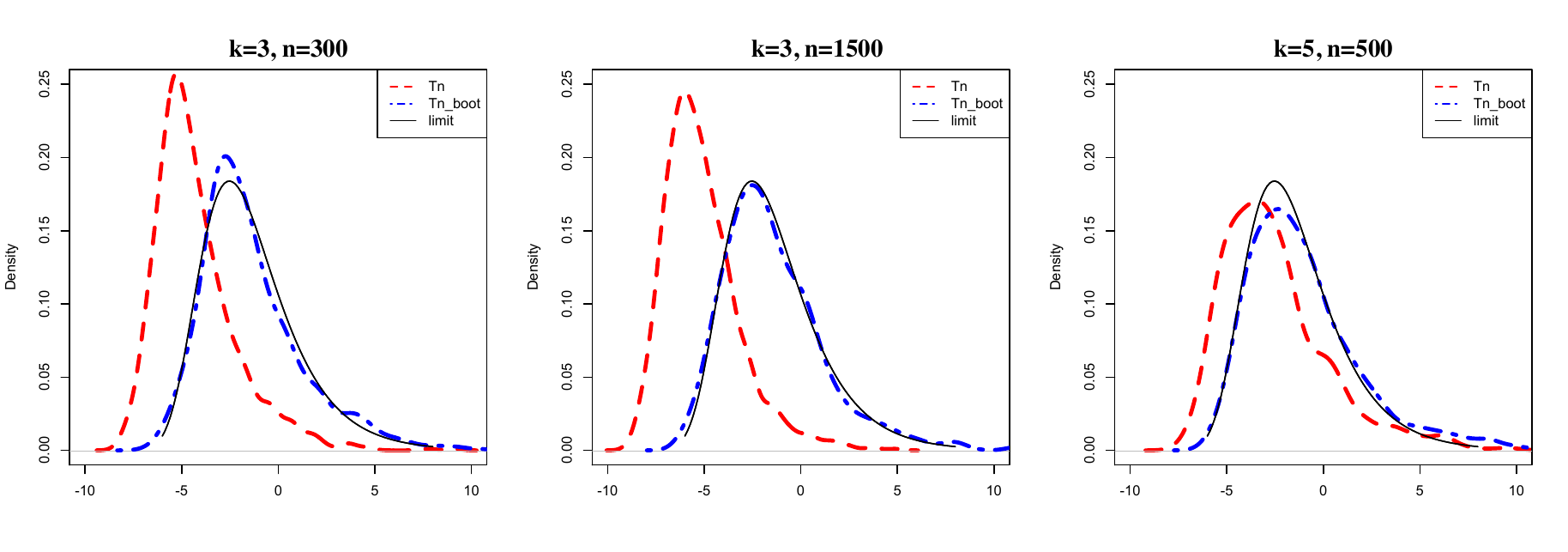}
\caption{Null densities under the degree-corrected stochastic block model setting in Simulation 4 with $k=3, n=300$ (left plot), $k=3, n=1500$ (middle plot) and $k=5, n=500$ (right plot). The red dashed lines, blue dash-dotted lines and black solid lines show the densities of the test statistic $T_{n2}$, the bootstrap corrected test statistic $T_{n2,\text{boot}}$ and the theoretical limit, respectively.}
\label{figure-b}
\end{figure}

In Figure \ref{figure-b}, we plot the distribution of $T_{n2}$ with and without bootstrap corrections from 1000 data replications.
It can be seen that when $k=3$, the null distribution of $T_{n2}$ deviates from the Gumbel distribution by a location and scale shift.
Such a deviation does not decrease even when the sample size is increased to $n=1500$.
However, when $k$ is increased to 5, the distribution of $T_{n2}$ is much less deviated from the Gumbel distribution in Theorem \ref{theorem:null:b}.
Note that when the bootstrap correction is considered, the sample null distribution is close to the limit even when $k=3$ and $n = 300$.}

\medskip
\noindent \textbf{Simulation 5. Hypothesis test under the degree-corrected stochastic block model.}
\noindent {In the degree-corrected stochastic block model setting, we then consider the hypothesis test
$$H_0: k=k_0\,\,\,vs\,\,\, H_1: k>k_0.$$
We investigate the probability of type I error and the power of the test statistic $T^+_{n2,\text{boot}}$, which is the augmented test statistic (calculated as in Section \ref{sec:augmentdcsbm}) with bootstrap correction.
The connecting probabilities are generated the same as in Simulation 4.
We consider equal sized communities with 200 nodes each.
Each simulation is repeated 100 times. The simulation results are given in Table \ref{tab:g}.
It is seen that the probability of type I error is close to the nominal level and the test also shows good power.
}
\begin{table}[!t]
\caption{Proportion of rejection at nominal level $\alpha=0.05$ for hypothesis test $H_0: k=k_0$ vs $H_1: k>k_0$ under the setting in Simulation 5.}
\vspace{0.15in}
\centering
\scalebox{0.75}{
\renewcommand\arraystretch{1.5}
\begin{tabular}{c|ccccc}
\hline
& \multicolumn{5}{c}{$T^+_{n,\text{boot}}$} \\
\hline
$k$                            &2 & 4 & 6 & 8 & 10\\ \hline
$k_0=2$                    &\textbf{0.05}&1.00&1.00&1.00&1.00\\
$k_0=4$                    &*&\textbf{0.06}&1.00&1.00&1.00\\
$k_0=6$                    &*&*&\textbf{0.04}&1.00&1.00\\
$k_0=8$                    &*&*&*&\textbf{0.04}&1.00\\
$k_0=10$                  &*&*&*&*&\textbf{0.07}   \\
\hline
\end{tabular}}\label{tab:g}
\end{table}

\section{Data examples}
\label{section:dataexample}

\subsection{International trade data}
In this subsection, we apply the proposed method to the international trade dataset that was studied in \cite{Westveld:Hoff:2011}.
The dataset contains yearly international trade information among $n=58$ countries from 1981--2000.  The original network is directed and weighted, in which each node corresponds to a country and for a given year, $\textrm{Trade}_{ij}$ indicates the amount of exports from country $i$ to country $j$; see \cite{Westveld:Hoff:2011} for details.  \cite{Saldana:Yu:Feng:2017} revisited the dataset for the purpose of estimating the number of communities.  Following \cite{Saldana:Yu:Feng:2017}, we focus on the international trade network in 1995 and transform the directed and weighted adjacency matrix to an undirected binary matrix.  Specifically, let $W_{ij}=\textrm{Trade}_{ij} + \textrm{Trade}_{ji}$, and we set $A_{ij}=1$ if $W_{ij}\geq W_{0.5}$, and $A_{ij}=0$ otherwise, where $W_{0.5}$ denotes the 50th percentile of $\{W_{ij}\}_{1\leq i<j\leq n}$.
\cite{Saldana:Yu:Feng:2017} used three different methods and identified three different numbers of communities, specifically, 3, 7 and 10 respectively.  {Note due to the limited size of the network, some communities can be very small (e.g., less than 5 nodes). In that case, the augmentation procedure may not work well since the added community may have less than or equal to 2 nodes.
We thus use $T_{n,\text{boot}}$ as the test statistic for the stochastic block model and obtain $T_{n,\text{boot}} =$ 44.28, 3.33 and 13.44 for $k_0=$3, 7 and 10, respectively. Since $t_{0.95}=3.41$ for the Gumbel distribution, we do not reject $H_0: k=7$ at the level of 0.05.}
As discussed in \cite{Saldana:Yu:Feng:2017}, $k=7$ seems to be a reasonable choice for the number of communities, corresponding to countries with highest GDPs, industrialized European and Asian countries with medium-level GDPs, and developing countries in South America with the lowest GDPs.

\subsection{Political blog data}
In this subsection, we use the political blog network \citep{Adamic:Glance:2005} to demonstrate the proposed methods for both the stochastic block model and the degree-corrected stochastic block model. The dataset consists of political blogs, with edges representing web links. Each node is labeled either as ``conservative'' or ``liberal'' based on the blogger's political stance. We only consider the largest connected component of this network which consists of $1222$ nodes as is commonly done in the literature.
\cite{Chen:Lei:2017} applied a network cross-validation method to the political blog data to select the number of communities, and they identified $k=10$ and $k=2$ respectively for the stochastic block model and the degree-corrected stochastic block model.
Here, we reanalyze the data using the test statistics that we have developed to test the significance of the number of communities identified by \cite{Chen:Lei:2017}.
{Specifically, we use $T^+_{n,\text{boot}}$ and $T^+_{n2,\text{boot}}$ as the test statistic for the stochastic block model and the degree-corrected stochastic block model, respectively.
We obtain $T^+_{n,\text{boot}}$=23.59 under $k=10$, and $T^+_{n2,\text{boot}}$=2.06 under $k=2$.
Since $t_{0.95}=3.41$ for the Gumbel distribution, at the level of 0.05, we would reject $H_0: k=10$ under the stochastic block model and not reject $H_0: k=2$ under the degree-corrected stochastic block model.
The result of our analysis agrees with that of \cite{Chen:Lei:2017} for the degree-corrected stochastic block model but not so for the stochastic block model. It is possible that the stochastic block model is not an appropriate model for this particular dataset as it was observed that there is a big variation among node degrees.
}
%

\section{Discussion}

In this article, we have developed a novel goodness-of-fit test based on the maximum entry-wise deviation of the centered and re-scaled observed adjacency matrix and demonstrated that its asymptotic null distribution is the Gumbel distribution when $k=o(n/\log^2n)$, which significantly relaxes the condition in \cite{Lei:2016}.
The test is different from those used in traditional methods based on independent random variables, in which the goodness-of-fit is assessed by the sum of residual squares.
For stochastic block models, the residual is a matrix, and the proposed test incorporates the signal change among different blocks nested in the residual matrix to test the goodness-of-fit of the model.
In the case of degree-corrected stochastic block model with unknown degree parameters, through simulation studies, we show that the distribution of $T_{n2}$ under the null deviates from the Gumbel distribution with $\mu=-2\log(2\sqrt{\pi})$ and $\beta=2$. Finding the asymptotic null distribution of $T_{n2}$ under the degree-corrected stochastic block model is a challenging task, as the estimated degree parameters $\widehat\omega_i$, $i=1,\ldots,n$, introduce complex dependencies between the entries of the re-scaled adjacency matrix.
As such, the theoretical arguments used in the current article can no longer be directly applied or extended to obtain the asymptotic null distribution of $T_{n2}$.

{Both our work and \cite{Lei:2016} consider dense networks, i.e., entries in the probability matrix $B$ are bounded away from 0. In practice, networks can be sparse. To admit the sparse case, the probability matrix $B$ is often assumed to be of the form $B = \rho_n B_0$, where the entries of $B_0$ are of order $1$, and $\rho_n$ is a parameter controlling the sparsity of the network and allowed to decrease to zero when $n$ increases \citep{Bickel:Chen:2009}.
Some recent work that consider tests based on subgraph statistics have obtained closed-form asymptotic null distributions in the sparse regime \citep{Gao:Lafferty:2017,Jin:Ke:Luo:2019}. However, these asymptotic null distributions only hold under $k=1$, i.e., the Erd\H{o}s--R\'{e}nyi model. Thus, these tests can only be used to test if there are communities in the network but not how many are there.
In addition, \cite{Wang:Bickel:2017} considers a likelihood-based model selection approach and derives the asymptotic distribution of the log-likelihood ratio statistic in the sparse case. However, the number of communities $k$ is fixed in their work.
For our method, when the network is sparse, existing arguments do not guarantee the moderate deviation bound (see Lemma \ref{lemma:chen}) due to the heavy tail of the centered and re-scaled adjacency entry $(A_{ij}-B_{\sigma(i)\sigma(j)})/\sqrt{B_{\sigma(i)\sigma(j)}(1-B_{\sigma(i)\sigma(j)})}$. It would be of interest to investigate whether the moderate deviation bound in Lemma \ref{lemma:chen} can be modified to consider such heavy-tailed scenarios in future work.

}

\newpage
\renewcommand{\thesubsection}{S\arabic{subsection}}
\renewcommand{\theequation}{S\arabic{equation}}
\renewcommand{\thetable}{S\arabic{table}}
\setcounter{equation}{0}
\setcounter{table}{0}
\setcounter{subsection}{0}
\setcounter{page}{1}
\def\eop
{\hfill $\Box$
}

\begin{center}
{\Large\bf Supplementary Materials} \\
{\large\bf Using Maximum Entry-Wise Deviation to Test the Goodness-of-Fit for Stochastic Block Models} \\
\medskip
\end{center}

\subsection{Proof of Theorem \ref{theorem:null:a}, Theorem \ref{theorem:alternative:a} and Corollary \ref{cor:alternative}}
We start with three lemmas that will be used in the proof.
The following Poisson approximation result is essentially a special case of Theorem \ref{theorem:null:a} in \cite{Arritia:Goldstein:Gordon:1989}.

\begin{lemma}[\cite{Arritia:Goldstein:Gordon:1989}]
\label{lemma:poisson}
Let $I$ be an index set and $\{\mathbf{B}_\alpha, \alpha\in I\}$ be a set of subsets of $I$, that is, $\mathbf{B}_\alpha\subset I$. Let also $\{\eta_\alpha, \alpha\in I\}$ be random variables. For a given $t\in \mathbb{R}$, set $\lambda=\sum_{\alpha\in I}P(\eta_\alpha>t)$. Then
$$\mid P(\max_{\alpha\in I} \eta_\alpha\leq t)-e^{-\lambda}\mid\leq (1\wedge\lambda^{-1})(b_1+b_2+b_3),$$
where
$$b_1=\sum_{\alpha\in I}\sum_{\beta\in \mathbf{B}_\alpha}P(\eta_\alpha>t)P(\eta_\beta>t),\,\,b_2=\sum_{\alpha\in I}\sum_{\alpha\neq\beta\in \mathbf{B}_\alpha}P(\eta_\alpha>t,\eta_\beta>t),$$
$$b_3=\sum_{\alpha\in I}E\mid P(\eta_\alpha>t|\sigma(\eta_\beta,\beta\notin \mathbf{B}_\alpha))-P(\eta_\beta>t)\mid,$$
and $\sigma(\eta_\beta,\beta\notin \mathbf{B}_\alpha)$ is the $\sigma$-algebra generated by $\{\eta_\beta,\beta\notin \mathbf{B}_\alpha\}$. In particular, if $\eta_\alpha$ is independent of $\{\eta_\beta,\beta\notin \mathbf{B}_\alpha\}$ for each $\alpha$, then $b_3=0$.
\end{lemma}

The following moderate deviation result is from \cite{Chen:1990}.

\begin{lemma}[\cite{Chen:1990}]
\label{lemma:chen}
Suppose $\xi_1,\xi_2,\ldots,\xi_n$ are i.i.d random variables with $E\xi_1=0$ and $E\xi_1^2=1$. Set $S_n=\sum_{i=1}^n\xi_i$. Let $0<\alpha\leq 1$ and $\{a_n:n\geq 1\}$ satisfy that $a_n\rightarrow\infty$ and $a_n=o(n^{\alpha/(2(2-\alpha))})$. If $Ee^{t_0|\xi_1|^\alpha}<\infty$ for some $t_0>0$, then
$$\lim_n\frac{1}{a_n^2}\log P(\frac{S_n}{\sqrt{n}a_n}\geq \mu)=-\frac{\mu^2}{2}$$
for any $\mu>0$.
\end{lemma}

The following result is  from \cite{Cai:Jiang:2011}.

\begin{lemma}[\cite{Cai:Jiang:2011}]\label{lemma:CJ}
Suppose $\xi_1,\xi_2,\ldots,\xi_n$ are i.i.d random variables with $E\xi_1=0$, $E\xi_1^2=1$ and $Ee^{t_0|\xi_1|^\alpha}<\infty$ for some $t_0>0$ and $0<\alpha\leq 1$. Set $S_n=\sum_{i=1}^n\xi_i$ and $\beta=\alpha/(2+\alpha)$. Then, for any $\{p_n:n\geq 1\}$ with  $0<p_n\rightarrow\infty$ and $\log p_n=o(n^\beta)$ and $\{y_n;n\geq 1\}$ with $y_n\rightarrow y>0$,
$$P(\frac{S_n}{\sqrt{n\log p_n}}\geq y_n)\sim \frac{p_n^{-y_n^2/2}(\log{p_n})^{-1/2}}{\sqrt{2\pi}y}$$
as $n\rightarrow\infty$.\\
\end{lemma}

\noindent {\bf Proof of Theorem \ref{theorem:null:a}.}
By \citeauthor{Hoeffding:1963}'s (\citeyear{Hoeffding:1963}) inequality, we have
\[
\begin{array}{lll}
P(\max_{1\leq u\leq v\leq k}\mid B_{uv}-\widehat{B}_{uv}\mid>t)& \leq & k^2e^{-2\mid\sigma^{-1}(u)\mid\mid\sigma^{-1}(v)\mid t^2}\\
& \leq & e^{2\log k-2\mid\sigma^{-1}(u)\mid\mid\sigma^{-1}(v)\mid t^2}\\
& \leq & e^{2\log k-2C_1^2n^2t^2/k^2}.
\end{array}
\]
It implies that
\[
\max_{1\leq i\leq j\leq n}\mid B_{\sigma(i)\sigma(j)}-\widehat{B}_{\sigma(i)\sigma(j)}\mid=o_p(\frac{ k\log n }{ n }).
\]
Let
\[
\begin{array}{lll}L_{n,0}
&\triangleq&\max_{1\leq i\leq n,1\leq v\leq k_0} \mid\widehat{\rho}_{iv,0}\mid\\
&=&\max_{1\leq i\leq n,1\leq v\leq k_0}\mid\frac{1}{\sqrt{\mid\sigma_0^{-1}(v)/\{i\}\mid}}\sum_{j\in\sigma_0^{-1}(v)/\{i\}}\frac{A_{ij}-B^{\sigma_0}_{\sigma_0(i)\sigma_0(j)}}{\sqrt{B_{\sigma(i)\sigma(j)}(1-B_{\sigma(i)\sigma(j)})}}\mid\\
&=&\max_{1\leq i\leq n,1\leq v\leq k}\mid\frac{1}{\sqrt{\mid\sigma^{-1}(v)/\{i\}\mid}}\sum_{j\in\sigma^{-1}(v)/\{i\}}\frac{A_{ij}-B_{\sigma(i)\sigma(j)}}{\sqrt{B_{\sigma(i)\sigma(j)}(1-B_{\sigma(i)\sigma(j)})}}\mid,\end{array}
\]

\[
\begin{array}{lll}L_{n,1}
&\triangleq&\max_{1\leq i\leq n,1\leq v\leq k_0} \mid\widehat{\rho}_{iv,1}\mid\\
&=&\max_{1\leq i\leq n,1\leq v\leq k_0}\mid\frac{1}{\sqrt{\mid\sigma_0^{-1}(v)/\{i\}\mid}}\sum_{j\in\sigma_0^{-1}(v)/\{i\}}\frac{A_{ij}-\widehat{B}^{\sigma_0}_{\sigma_0(i)\sigma_0(j)}}{\sqrt{B_{\sigma(i)\sigma(j)}(1-B_{\sigma(i)\sigma(j)})}}\mid\\
&=&\max_{1\leq i\leq n,1\leq v\leq k_0}\mid\widehat{\rho}_{iv,0}+\frac{1}{\sqrt{\mid\sigma_0^{-1}(v)/\{i\}\mid}}\sum_{j\in\sigma^{-1}(v)/\{i\}}\frac{B^{\sigma_0}_{\sigma_0(i)\sigma_0(j)}-\widehat{B}^{\sigma_0}_{\sigma_0(i)\sigma_0(j)}}{\sqrt{B_{\sigma(i)\sigma(j)}(1-B_{\sigma(i)\sigma(j)})}}\mid\\
&=&\max_{1\leq i\leq n,1\leq v\leq k}\mid\widehat{\rho}_{iv,0}\mid+o_p(\sqrt{\mid\sigma^{-1}(v)\mid}\frac{ k\log n }{ n })\\
&=&L_{n,0}+o_P(1),
\end{array}
\]

\[
\begin{array}{lll}L_{n,2}
&\triangleq&\max_{1\leq i\leq n,1\leq v\leq k_0} \mid\widehat{\rho}_{iv,2}\mid\\
&=&\max_{1\leq i\leq n,1\leq v\leq k_0}\mid\frac{1}{\sqrt{\mid\sigma_0^{-1}(v)/\{i\}\mid}}\sum_{j\in\sigma_0^{-1}(v)/\{i\}}\frac{A_{ij}-\widehat{B}^{\sigma_0}_{\sigma_0(i)\sigma_0(j)}}{\sqrt{\widehat{B}^{\sigma_0}_{\sigma_0(i)\sigma_0(j)}(1-\widehat{B}^{\sigma_0}_{\sigma_0(i)\sigma_0(j)})}}\mid\\
&=&\max_{1\leq i\leq n,1\leq v\leq k_0} (\mid\frac{1}{\sqrt{\mid\sigma_0^{-1}(v)/\{i\}\mid}}\sum_{j\in\sigma_0^{-1}(v)/\{i\}}\frac{A_{ij}-\widehat{B}^{\sigma_0}_{\sigma_0(i)\sigma_0(j)}}{\sqrt{B_{\sigma(i)\sigma(j)}(1-B_{\sigma(i)\sigma(j)})}}\frac{\sqrt{B_{\sigma(i)\sigma(j)}(1-B_{\sigma(i)\sigma(j)})}}{\sqrt{\widehat{B}^{\sigma_0}_{\sigma_0(i)\sigma_0(j)}(1-\widehat{B}^{\sigma_0}_{\sigma_0(i)\sigma_0(j)})}}\mid\\
&=&L_{n,1}(1+o_P(\sqrt{\frac{ k\log n }{ n }}))\\
&=&L_{n,0}+L_{n,0}o_P(\sqrt{\frac{ k\log n }{ n }})+o_P(1).
\end{array}
\]

If $k=o(n/\log^2 n)$ and $L_{n,0}=O_p(\sqrt{\log n})$, we have
\[
L_{n,2}=L_{n,0}+o_P(1).
\]

Thus, to prove Theorem \ref{theorem:null:a} (\ref{eq:gumbel1}), it is sufficient to show:
\[
\frac{L_{n,0}}{\sqrt{\log(2kn)}}\rightarrow \sqrt{2}
\]
in probability as $n\rightarrow\infty$.

We first prove
\[
\lim_{n\to\infty} P(\frac{L_{n,0}}{\sqrt{\log(2kn)}}\leq \sqrt{2}-\epsilon)=0,
\]
for any $\epsilon>0$ small enough.

Let $y_n=(\sqrt{2}-\epsilon)\sqrt{\log(2kn)}$, $I=\{(i,v)|1\leq i\leq n,1\leq v\leq k\}$, $\mathbf{B}_{iv}=\{(s,t)\in I/\{(i,v)\}|s=i\}$. Then $\mid\mathbf{B}_{iv}\mid=k-1$.
Note that $E(\frac{A_{ij}-B_{\sigma(i)\sigma(j)}}{\sqrt{B_{\sigma(i)\sigma(j)}(1-B_{\sigma(i)\sigma(j)})}})=0$ and $E(\frac{A_{ij}-B_{\sigma(i)\sigma(j)}}{\sqrt{B_{\sigma(i)\sigma(j)}(1-B_{\sigma(i)\sigma(j)})}})^2=1$. By
Lemma \ref{lemma:poisson}, we have $$\mid P(\max_{1\leq i\leq n,1\leq v\leq k}|\widehat{\rho}_{iv,0}|\leq y_n)-e^{-\lambda_n}\mid\leq b_1+b_2,$$
where $\lambda_n=\sum_{1\leq i\leq n,1\leq v\leq k}P(|\widehat{\rho}_{iv,0}|>y_n)$.
By Lemma \ref{lemma:chen}, we have
$$\begin{array}{lll}\lambda_n&=&\sum_{1\leq i\leq n,1\leq v\leq k}P(|\widehat{\rho}_{iv,0}|>y_n)\\
&\leq&2kne^{-(\sqrt{2}-\epsilon)^2\log(2kn)/2}\\
&=&e^{\log(2kn)-(\sqrt{2}-\epsilon)^2\log(2kn)/2}\\
&\rightarrow&\infty,
\end{array}$$

$$\begin{array}{lll}b_1&=&\sum_{\alpha\in I}\sum_{\beta\in \mathbf{B}_\alpha}P(\eta_\alpha>y_n)P(\eta_\beta>y_n)\\
&<&4k^2ne^{-(\sqrt{2}-\epsilon)^2\log(2kn)}\\
&=&e^{\log(4k^2n)-(\sqrt{2}-\epsilon)^2\log(2kn)}\\
&=&o(1),
\end{array}$$

$$\begin{array}{lll}b_2&=&\sum_{\alpha\in I}\sum_{\alpha\neq\beta\in \mathbf{B}_\alpha}P(\eta_\alpha>y_n,\eta_\beta>y_n))\\
&<&4k^2ne^{-(\sqrt{2}-\epsilon)^2\log(2kn)}\\
&=&e^{\log(4k^2n)-(\sqrt{2}-\epsilon)^2\log(2kn)}\\
&=&o(1)
\end{array}$$
for sufficiently large $n$.
To finish the proof, we only need to show that for any $\epsilon>0$,
\[
\lim_{n\to\infty} P(\frac{L_{n,0}}{\sqrt{\log(2kn)}}\geq \sqrt{2}+\epsilon)=0.
\]
By Lemma \ref{lemma:chen}, we have
$$\begin{array}{lll}P(\frac{L_{n,0}}{\sqrt{\log(2kn)}}\geq \sqrt{2}+\epsilon)&=&P(\max_{1\leq i\leq n,1\leq v\leq k} \mid\widehat{\rho}_{iv,0}\mid\geq \sqrt{2}+\epsilon)\\
&\leq&2kne^{-(\sqrt{2}+\epsilon)^2\log(2kn)/2}\\
&=&e^{\log(2kn)-(\sqrt{2}+\epsilon)^2\log(2kn)/2}\\
&=&o(1)
\end{array}$$
for sufficiently large $n$.

Next, we show the second part (\ref{eq:gumbel2}) of Theorem \ref{theorem:null:a}.
Similar to the proof of Theorem \ref{theorem:null:a} (2.1), it is sufficient to show
\[
\lim_{n}P(L_{n,0}^2-2\log(2kn)+\mathrm{\log\log}(2kn)\leq y)=\exp\{-\frac{1}{2\sqrt{\pi}}e^{-y/2}\}.
\]

Let $y_n=\sqrt{y+2\log(2kn)-\log\log(2kn)}$, $I=\{(i,v)|1\leq i\leq n,1\leq v\leq k\}$, $\mathbf{B}_{iv}=\{(s,t)\in I/\{(i,v)\}|s=i\}$. Then $\mid\mathbf{B}_{iv}\mid=k-1$.
Note that $E(\frac{A_{ij}-B_{\sigma(i)\sigma(j)}}{\sqrt{B_{\sigma(i)\sigma(j)}(1-B_{\sigma(i)\sigma(j)})}})=0$ and $E(\frac{A_{ij}-B_{\sigma(i)\sigma(j)}}{\sqrt{B_{\sigma(i)\sigma(j)}(1-B_{\sigma(i)\sigma(j)})}})^2=1$. By
Lemma \ref{lemma:poisson}, we have $$\mid P(\max_{1\leq i\leq n,1\leq v\leq k}|\widehat{\rho}_{iv,0}|\leq y_n)-e^{-\lambda_n}\mid\leq b_1+b_2,$$
where $\lambda_n=\sum_{1\leq i\leq n,1\leq v\leq k}P(|\widehat{\rho}_{iv,0}|>y_n)$. By Lemma \ref{lemma:CJ}, we have
$$\begin{array}{lll}P(\widehat{\rho}_{iv,0}>y_n)\\
=P(\frac{\widehat{\rho}_{iv,0}}{\sqrt{\log(2kn)}}>\sqrt{\frac{y+2\log(2kn)-\mathrm{\log\log}(2kn)}{\log(2kn)}})\\
\sim(2kn)^{-\frac{y+2\log(2kn)-\log\log(2kn)}{2\log(2kn)}}(\log(2kn))^{-\frac{1}{2}}/(2\sqrt{\pi})\\
=(2kn)^{-1}(2kn)^{-\frac{y}{2\log(2kn)}}(2kn)^{\frac{\log\log(2kn)}{2\log(2kn)}}(\log(2kn))^{-\frac{1}{2}}/(2\sqrt{\pi})\\
=(2kn)^{-1}e^{-\frac{y}{2\log(2kn)}\log(2kn)}e^{\frac{\log\log(2kn)}{2\log(2kn)}\log(2kn)}(\log(2kn))^{-\frac{1}{2}}/(2\sqrt{\pi})\\
=(2kn)^{-1}e^{-\frac{y}{2}}e^{\log((\log(2kn))^{\frac{1}{2}}}(\log(2kn))^{-\frac{1}{2}}/(2\sqrt{\pi})\\
=(2kn)^{-1}e^{-\frac{y}{2}}(\log(2kn))^{\frac{1}{2}}(\log(2kn))^{-\frac{1}{2}}/(2\sqrt{\pi})\\
=(2kn)^{-1}e^{-\frac{y}{2}}/(2\sqrt{\pi}).
\end{array}$$
Hence,
$$\begin{array}{lll}\lambda_n&=&\sum_{1\leq i\leq n,1\leq v\leq k}P(|\widehat{\rho}_{iv,0}|>y_n)\\
&=&kn\frac{(kn)^{-1}}{2\sqrt{\pi}}e^{-\frac{y}{2}}\\
&=&\frac{1}{2\sqrt{\pi}}e^{-\frac{y}{2}}.
\end{array}$$
Similar to the proof for Theorem \ref{theorem:null:a} (\ref{eq:gumbel1}), we have
$b_1=o(1)$, $b_2=o(1)$.
Thus, $$\lim_nP(L_{n,0}\leq y_n)=\exp\{-\frac{1}{2\sqrt{\pi}}e^{-\frac{y}{2}}\}.$$\qed\\

\noindent {\bf Proof of Theorem \ref{theorem:alternative:a}.}
By \citeauthor{Hoeffding:1963}'s (\citeyear{Hoeffding:1963}) inequality, we have
\[
\begin{array}{lll}
P(\max_{1\leq u\leq v\leq k_0}\mid B_{uv}^{\sigma_0}-\widehat{B}_{uv}^{\sigma_0}\mid>t)& = &\sum_{1\leq u\leq v\leq k_0}P(\mid B_{uv}^{\sigma_0}-\widehat{B}_{uv}^{\sigma_0}\mid>t)\\
& \leq & k_0^2e^{-2\mid\sigma_0^{-1}(u)\mid\mid\sigma_0^{-1}(v)\mid t^2}\\
& \leq & e^{2\log k_0-2C_1^2n^2t^2/k_0^2}.
\end{array}
\]
It implies that
\[
\max_{1\leq i\leq j\leq n}\mid B^{\sigma_0}_{\sigma_0(i)\sigma_0(j)}-\widehat{B}^{\sigma_0}_{\sigma_0(i)\sigma_0(j)}\mid=o_p(\frac{ k_0\log n }{ n }).
\]

Let
\[
\begin{array}{lll}L_{n,0}
&\triangleq&\max_{1\leq i\leq n,1\leq v\leq k_0} \mid\widehat{\rho}_{iv,0}\mid\\
&=&\max_{1\leq i\leq n,1\leq v\leq k_0}\mid\frac{1}{\sqrt{\mid\sigma_0^{-1}(v)/\{i\}\mid}}\sum_{j\in\sigma_0^{-1}(v)/\{i\}}\frac{A_{ij}-B_{\sigma(i)\sigma(j)}}{\sqrt{B^{\sigma_0}_{\sigma_0(i)\sigma_0(j)}(1-B^{\sigma_0}_{\sigma_0(i)\sigma_0(j)})}}\mid,\end{array}
\]
\[
\begin{array}{lll}L_{n,1}
&\triangleq&\max_{1\leq i\leq n,1\leq v\leq k_0} \mid\widehat{\rho}_{iv,1}\mid\\
&=&\max_{1\leq i\leq n,1\leq v\leq k_0}\mid\frac{1}{\sqrt{\mid\sigma_0^{-1}(v)/\{i\}\mid}}\sum_{j\in\sigma_0^{-1}(v)/\{i\}}\frac{A_{ij}-\widehat{B}^{\sigma_0}_{\sigma_0(i)\sigma_0(j)}}{\sqrt{B^{\sigma_0}_{\sigma_0(i)\sigma_0(j)}(1-B^{\sigma_0}_{\sigma_0(i)\sigma_0(j)})}}\mid\\
&=&\max_{1\leq i\leq n,1\leq v\leq k_0}\mid\widehat{\rho}_{iv,0}+\frac{1}{\sqrt{\mid\sigma_0^{-1}(v)/\{i\}\mid}}\sum_{j\in\sigma_0^{-1}(v)/\{i\}}\frac{B_{\sigma(i)\sigma(j)}-\widehat{B}^{\sigma_0}_{\sigma_0(i)\sigma_0(j)}}{\sqrt{B^{\sigma_0}_{\sigma_0(i)\sigma_0(j)}(1-B^{\sigma_0}_{\sigma_0(i)\sigma_0(j)})}}\mid\\
&\geq&\max_{1\leq i\leq n,1\leq v\leq k_0}\mid\frac{1}{\sqrt{\mid\sigma_0^{-1}(v)/\{i\}\mid}}\sum_{j\in\sigma_0^{-1}(v)/\{i\}}\frac{B_{\sigma(i)\sigma(j)}-\widehat{B}^{\sigma_0}_{\sigma_0(i)\sigma_0(j)}}{\sqrt{B^{\sigma_0}_{\sigma_0(i)\sigma_0(j)}(1-B^{\sigma_0}_{\sigma_0(i)\sigma_0(j)})}}\mid-\max_{1\leq i\leq n,1\leq v\leq k_0}\mid\widehat{\rho}_{iv,0}\mid\\
&\geq&\max_{1\leq i\leq n,1\leq v\leq k_0}\mid\frac{1}{\sqrt{\mid\sigma_0^{-1}(v)/\{i\}\mid}}\sum_{j\in\sigma_0^{-1}(v)/\{i\}}\frac{B_{\sigma(i)\sigma(j)}-B^{\sigma_0}_{\sigma_0(i)\sigma_0(j)}}{\sqrt{B^{\sigma_0}_{\sigma_0(i)\sigma_0(j)}(1-B^{\sigma_0}_{\sigma_0(i)\sigma_0(j)})}}\mid\\
&&-\max_{1\leq i\leq n,1\leq v\leq k_0}\mid\frac{1}{\sqrt{\mid\sigma_0^{-1}(v)/\{i\}\mid}}\sum_{j\in\sigma_0^{-1}(v)/\{i\}}\frac{B^{\sigma_0}_{\sigma_0(i)\sigma_0(j)}-\widehat{B}^{\sigma_0}_{\sigma_0(i)\sigma_0(j)}}{\sqrt{B^{\sigma_0}_{\sigma_0(i)\sigma_0(j)}(1-B^{\sigma_0}_{\sigma_0(i)\sigma_0(j)})}}\mid-\max_{1\leq i\leq n,1\leq v\leq k_0}\mid\widehat{\rho}_{iv,0}\mid\\
&=&\max_{1\leq i\leq n,1\leq v\leq k_0}\mid\frac{1}{\sqrt{\mid\sigma_0^{-1}(v)/\{i\}\mid}}\sum_{j\in\sigma_0^{-1}(v)/\{i\}}\frac{B_{\sigma(i)\sigma(j)}-B^{\sigma_0}_{\sigma_0(i)\sigma_0(j)}}{\sqrt{B^{\sigma_0}_{\sigma_0(i)\sigma_0(j)}(1-B^{\sigma_0}_{\sigma_0(i)\sigma_0(j)})}}\mid\\
&&+o_p(\sqrt{\mid\sigma_0^{-1}(v)\mid}\frac{ k_0\log n }{ n })-\max_{1\leq i\leq n,1\leq v\leq k_0}\mid\widehat{\rho}_{iv,0}\mid\\
&=&\ell(k_0,\sigma_0)-L_{n,0}+o_P(1),
\end{array}
\]
\[
\begin{array}{lll}L_{n,2}
&\triangleq&\max_{1\leq i\leq n,1\leq v\leq k_0} \mid\widehat{\rho}_{iv,2}\mid\\
&=&\max_{1\leq i\leq n,1\leq v\leq k_0}\mid\frac{1}{\sqrt{\mid\sigma_0^{-1}(v)/\{i\}\mid}}\sum_{j\in\sigma_0^{-1}(v)/\{i\}}\frac{A_{ij}-\widehat{B}^{\sigma_0}_{\sigma_0(i)\sigma_0(j)}}{\sqrt{\widehat{B}^{\sigma_0}_{\sigma_0(i)\sigma_0(j)}(1-\widehat{B}^{\sigma_0}_{\sigma_0(i)\sigma_0(j)})}}\mid\\
&=&\max_{1\leq i\leq n,1\leq v\leq k_0}\mid\frac{1}{\sqrt{\mid\sigma_0^{-1}(v)/\{i\}\mid}}\sum_{j\in\sigma_0^{-1}(v)/\{i\}}\frac{A_{ij}-\widehat{B}^{\sigma_0}_{\sigma_0(i)\sigma_0(j)}}{\sqrt{B^{\sigma_0}_{\sigma_0(i)\sigma_0(j)}(1-B^{\sigma_0}_{\sigma_0(i)\sigma_0(j)})}}\frac{\sqrt{B^{\sigma_0}_{\sigma_0(i)\sigma_0(j)}(1-B^{\sigma_0}_{\sigma_0(i)\sigma_0(j)})}}{\sqrt{\widehat{B}^{\sigma_0}_{\sigma_0(i)\sigma_)(j)}(1-\widehat{B}^{\sigma_0}_{\sigma_0(i)\sigma_0(j)})}})\mid\\
&=& L_{n,1}(1+O_p(\frac{ k_0\log n }{ n }))\\
&=&L_{n,1}(1+o_P(1)).
\end{array}
\]
By \citeauthor{Hoeffding:1963}'s (\citeyear{Hoeffding:1963}) inequality, we have
\[
\begin{array}{lll}
P(\max_{1\leq i\leq n,1\leq v\leq k_0}\mid\frac{1}{\sqrt{\mid\sigma_0^{-1}(v)/\{i\}\mid}}\sum_{j\in\sigma_0^{-1}(v)/\{i\}}\frac{A_{ij}-B_{\sigma(i)\sigma(j)}}{\sqrt{B^{\sigma_0}_{\sigma_0(i)\sigma_0(j)}(1-B^{\sigma_0}_{\sigma_0(i)\sigma_0(j)})}}\mid>t)\\
 \leq \sum_{1\leq i\leq n,1\leq v\leq k_0}P(\mid\sum_{j\in\sigma_0^{-1}(v)/\{i\}}(A_{ij}-B_{\sigma(i)\sigma(j)})\mid>t\sqrt{\mid\sigma_0^{-1}(v)/\{i\}\mid}\sqrt{B^{\sigma_0}_{\sigma_0(i)\sigma_0(j)}(1-B^{\sigma_0}_{\sigma_0(i)\sigma_0(j)})})\\
 \leq  2e^{\log (k_0n)-2C_1^2t^2}.
\end{array}
\]
 Hence,
 \[
 L_{n,0}=O_p(\sqrt{\log (k_0n)})=O_p(\sqrt{\log n}).
 \]
Let $T_n=L_n^2(k_0, \sigma_0) - 2\log(2k_0n) + \log\log(2k_0n)$. Since $(k_0,\sigma_0)\in\mathcal{F}_\gamma(k,\sigma,B)$, as $n\rightarrow\infty$, we have
\[
T_n/\sqrt{\log n}\rightarrow\infty.
\]
Thus, (\ref{eq:gumbel3}) holds.
\qed\\

\noindent {
{\bf Proof of Corollary \ref{cor:alternative}.}
Consider the SBM with $k,\sigma$ and $B$. Under $k_0<k$ and $\sigma_0$ satisfying $\sigma_0(i)=\sigma_0(j)$ if $\sigma(i)=\sigma(j)$, one community in $\sigma_0$ will contain at least two communities in $\sigma$.
We first consider the case of $k_0=k-1$.
In this case, one community in $\sigma_0$ contains exactly two communities in $\sigma$.
Without loss of generality, let community $\sigma^{-1}(1)$ and $\sigma^{-1}(2)$ be merged into $\sigma_0^{-1}(1)$. Let $n_u\triangleq|\sigma^{-1}(u)|$, $1\le u\le k$, and assume that $n_1\ge n_2$.
Define
\[
\begin{array}{lll}\ell_{iu}&\triangleq&\frac{1}{\sqrt{\mid\sigma_0^{-1}(u)/\{i\}\mid}}\sum_{j\in\sigma_0^{-1}(u)/\{i\}}(B_{\sigma(i)\sigma(j)}-B^{\sigma_0}_{\sigma_0(i)\sigma_0(j)}).
\end{array}
\]
Under this case, we have $\ell_{iu}=0$ for $i\notin\sigma_0^{-1}(1)$ and $u\neq 1$.
When $i\in\sigma^{-1}(1)$, we have
\[
\begin{array}{lll}\ell_{i1}&=&\frac{1}{\sqrt{\mid\sigma_0^{-1}(1)/\{i\}\mid}}\sum_{j\in\sigma_0^{-1}(1)/\{i\}}(B_{\sigma(i)\sigma(j)}-B^{\sigma_0}_{\sigma_0(i)\sigma_0(j)})\\
&=&\frac{1}{\sqrt{n_1+n_2-1}}\mid (n_1-1)B_{11}+n_2B_{12}-\frac{n_1(n_1-1)B_{11}+n_2(n_2-1)B_{22}+2n_1n_2B_{12}}{n_1+n_2}\mid\\
&=&\frac{1}{\sqrt{n_1+n_2}}\mid\frac{(n_1-1)n_2(B_{11}-B_{12})-n_2(n_2-1)(B_{22}-B_{12})}{n_1+n_2}\mid.
\end{array}
\]
Similarly, we can derive that when $i\in\sigma^{-1}(2)$, we have
\[
\begin{array}{lll}\ell_{i1}&=&\frac{1}{\sqrt{\mid\sigma_0^{-1}(1)/\{i\}\mid}}\sum_{j\in\sigma_0^{-1}(1)/\{i\}}(B_{\sigma(i)\sigma(j)}-B^{\sigma_0}_{\sigma_0(i)\sigma_0(j)})\\
&=&\frac{1}{\sqrt{n_1+n_2}}\mid\frac{n_1(n_2-1)(B_{11}-B_{22})-n_1(n_1-1)(B_{11}-B_{12})}{n_1+n_2}\mid.
\end{array}
\]
We can see that if $B_{11}\neq B_{12}$, we have that $\ell(k_0,\sigma_0)=\max_{iu}\ell_{iu}=O(|\sigma^{-1}(1)|)$.
If $B_{11}=B_{12}$ and $|n_1/n_2-1|\ge c_0$, we have that $\ell(k_0,\sigma_0)=\max_{iu}\ell_{iu}=O(|\sigma^{-1}(1)|)$.
The above arguments can be generalized to $k_0<k-1$ and here we omit the details.

\subsection{Additional Computational Details}
\subsubsection{Augmented test statistic under the DCSBM}\label{sec:augmentdcsbm}
For the degree-corrected stochastic block model, when carrying out hypothesis test (1), an additional community can be added to the observed network to improve the power of our proposed test.
Denote $k_0^+=k_0+1$.
For a given adjacency matrix $A$ and null hypothesis $H_0:\,k=k_0,\sigma=\sigma_0$, the augmented test statistic is calculated through the following steps:
\begin{enumerate}[(1)]
\item Calculate $\widehat B$ using \eqref{eq:bhat} and $\widehat\omega$ using its maximum likelihood estimate.
\item Add a $k_0^+$th community of size $n_{k_0^+}=\min_{1\le u\le k_0} |\sigma_0^{-1}(u)|/2$ to the observed network.
For the added community, let the within and between community connecting probabilities be $\max_{1\le u\le k_0}\widehat B_{uu}$ and $\min_{u\neq v}\widehat B_{uv}$, respectively. Let the nodes in the added community have degree parameters equal to 1.
\item Calculate the size $n^+$ and the adjacency matrix $A^+$ of the network from step (2). With $\sigma^+_0=(\sigma_0,\underbrace{k_0^+,\ldots,k_0^+}_{n_{k_0^+}})$, calculate $\widehat B^+$ and $\widehat\omega^+$.
\item The augmented test statistic is calculated as
$$
T^+_{n2}= L_{n2}^2(k_0^+, \sigma^+_0) - 2\log(2k_0^+n^+) + \log\log(2k_0^+n^+),
$$
where $ L_{n2}$ is defined in \eqref{eq:dcsbm}.
\end{enumerate}
To carry out hypothesis test (1), we reject $H_0: k=k_0$, if $T^+_{n2}>t_{(1-\alpha)}$, where $t_{\alpha}$ is the $\alpha$-th quantile of the Gumbel distribution with $\mu=-2\log(2\sqrt{\pi})$ and $\beta=2$.

\subsubsection{Bootstrap corrected test statistic under the DCSBM}\label{sec:bootdcsbm}
For an adjacency matrix $A$ and null hypothesis $k=k_0,\sigma=\sigma_0$, the bootstrap corrected goodness-of-fit test statistic under the degree-corrected stochastic block model is calculated as follows:
\begin{enumerate}
\item Calculate $\widehat B$ using \eqref{eq:bhat} and $\widehat\omega$ using its maximum likelihood estimate. Calculate $T_{n2}$ using $A$, $(\widehat B,\widehat\omega,\sigma_0)$.
\item For $m=1,\ldots,M$, generate $A^{(m)}$ from the degree-corrected stochastic block model $(\widehat B,\widehat\omega,\sigma_0)$, and calculate $T_{n2}^{(m)}$ using $A^{(m)}$ and $(\widehat B,\widehat\omega^{(i)},\sigma_0)$, where $\widehat\omega^{(m)}$ is the degree parameter calculated with $A^{(m)}$ and $\sigma_0$.
\item Using $(T_{n2}^{(m)}:1\le m\le M)$, estimate the location and scale parameters  $\widehat\mu_2$ and $\widehat\beta_2$ of the Gumbel distribution using maximum likelihood.
\item The bootstrap corrected test statistic is calculated as
$$
T_{n2,\text{boot}}=\mu+\beta\left(\frac{T_{n2}-\widehat\mu_2}{\widehat \beta_2}\right),
$$
where $\mu=-2\log(2\sqrt{\pi})$ and $\beta=2$.
\end{enumerate}
In all of our simulations, we use $M=100$.

\subsection{Additional Simulation Results}
\subsubsection{Hypothesis test (1) with varying sparsity, community sizes and random $B$}
\begin{table}[!t]
\caption{Proportion of rejection at nominal level $\alpha=0.05$ for hypothesis test $H_0: k=k_0$ vs $H_1: k>k_0$.}
\vspace{0.15in}
\centering
\scalebox{0.75}{
\renewcommand\arraystretch{1.5}
\begin{tabular}{c|c|c|cccc|cccc}
\hline
\multicolumn{1}{c}{} & \multicolumn{2}{c}{} & \multicolumn{4}{|c|}{$T^+_{n,\text{boot}}$} & \multicolumn{4}{c}{$\text{Lei}_{\text{boot}}$} \\\hline
\multirow{12}{*}{\begin{tabular}[c]{@{}c@{}}Planted \\ Partition \end{tabular}}
&\multirow{5}{*}{\begin{tabular}[c]{@{}c@{}}$\rho=0.05$\\ $n_1=n_2=300$ \end{tabular}}
&$k$                               & 2 & 5 & 10 &15                             & 2 & 5& 10 &15  \\\cline{2-11}
&&$k_0=2$                    &\textbf{0.05}&1.00&1.00&1.00&              \textbf{0.08}&1.00&1.00&1.00\\
&&$k_0=5$                    &*&\textbf{0.04}&1.00&1.00&                   *&\textbf{0.08}&1.00&1.00\\
&&$k_0=10$                  &*&*&\textbf{0.06}&0.97&                         *&*&\textbf{0.20}&1.00\\
&&$k_0=15$                  &*&*&*&\textbf{0.06}&                               *&*&*&\textbf{0.60}\\\cline{2-11}
&\multirow{4}{*}{\begin{tabular}[c]{@{}c@{}}$\rho=0.05$\\ $n_1=200$, $n_2=400$ \end{tabular}}
&$k_0=2$                      &\textbf{0.05}&0.99&1.00&1.00&              \textbf{0.04}&1.00&1.00&1.00\\
&&$k_0=5$                    &*&\textbf{0.07}&1.00&1.00&                   *&\textbf{0.10}&1.00&1.00\\
&&$k_0=10$                  &*&*&\textbf{0.06}&1.00&                         *&*&\textbf{0.26}&1.00\\
&&$k_0=15$                  &*&*&*&\textbf{0.08}&                               *&*&*&\textbf{0.78}\\\cline{2-11}
&\multirow{4}{*}{\begin{tabular}[c]{@{}c@{}}$\rho=0.025$\\ $n_1=n_2=300$ \end{tabular}}
&$k_0=2$                      &\textbf{0.08}&0.91&1.00&1.00&             \textbf{0.39}&1.00&1.00&1.00\\
&&$k_0=5$                    &*&\textbf{0.09}&0.99&1.00&                   *&\textbf{0.37}&1.00&1.00\\
&&$k_0=10$                  &*&*&\textbf{0.10}&0.99&                         *&*&\textbf{0.47}&1.00\\
&&$k_0=15$                  &*&*&*&\textbf{0.07}&                               *&*&*&\textbf{0.78}\\\hline
\multirow{4}{*}{Random $B$}
&\multirow{4}{*}{$n_1=n_2=200$}
&$k_0=2$                      &\textbf{0.05}&1.00&1.00&1.00&              \textbf{0.05}&1.00&1.00&1.00\\
&&$k_0=5$                    &*&\textbf{0.03}&1.00&1.00&                   *&\textbf{0.02}&1.00&1.00\\
&&$k_0=10$                  &*&*&\textbf{0.04}&0.88&                         *&*&\textbf{0.16}&1.00\\
&&$k_0=15$                  &*&*&*&\textbf{0.10}&                               *&*&*&\textbf{0.84}\\
\hline
\end{tabular}}\label{tab:vary}
\end{table}
We consider the stochastic block model with $B_{uv}=\rho(1+4\times\mathbf{1}(u=v))$, $\ceil{k/2}$ communities of size $n_1$ and $k-\ceil{k/2}$ communities of size $n_2$, where $\ceil{x}$ denotes the least integer greater than or equal to $x$.  We let the sparsity level $\rho=0.025, 0.05$, and community sizes $n_1=200,300$, $n_2=300,400$. We also consider the case where the entries in $B$ are randomly generated with $B_{uu}$ from Uniform $[0.2,0.5]$ and $B_{uv}$, $u\neq v$, from Uniform $[0.025, 0.1]$.
Table \ref{tab:vary} reports the results from $T^+_{n, \text{boot}}$ and Lei$_{n, \text{boot}}$ from 100 data replications.
For all settings, our test has type I errors close to the nominal level.
It is seen that \cite{Lei:2016} does not perform well when $k$ is large or when the graph is very sparse (i.e., $\rho=0.025$).
}

\subsubsection{Hypothesis test (2) under the stochastic block model}
\begin{table}[!t]
\caption{Proportion of rejection at nominal level $\alpha=0.05$ for hypothesis test $H_0: \sigma=\sigma_0$.}
\vspace{0.15in}
\label{tab:4a}
\centering
\scalebox{1}{
\begin{tabular}{cc|cc|cc}
\hline                                        &      & \multicolumn{2}{|c}{$n/k=100$}   & \multicolumn{2}{|c}{$n/k=200$} \\ \cline{3-6}
                                            &  $z\%$   & $r=0.05$  & $r=0.10$ & $r=0.05$         & $r=0.10$         \\ \hline
\multicolumn{1}{l}{}                       & 0.01  &0.80  & 1.00 &    1.00& 1.00            \\
\multicolumn{1}{l}{}                       & 0.05  & 0.96  & 1.00 &     1.00 & 1.00          \\
\multicolumn{1}{l}{}                       & 0.10 &  1.00  & 1.00 &     1.00 & 1.00    \\ \hline
\end{tabular}
}
\end{table}
We investigate the power of our proposed test for testing $H_0: \sigma=\sigma_0$ vs $H_1: \sigma\neq\sigma_0$.
Specifically, we investigate as we move $\sigma_0$ away from the true community assignment vector, if and when the test $H_0: \sigma=\sigma_0$ would be rejected at the nominal level. We consider stochastic block models with two equal-sized blocks with block size $100$ or $200$. The edge probability between communities $u$ and $v$ is $r(1+2\times\mathbf{1}(u=v))$, where $r=0.05$ or $0.10$.
We perform hypotheses tests with $\sigma_0=\sigma_z$, where $\sigma_z$ is the true community assignment vector with $z\%$ of the entries corrupted. We consider $z\%=0.01$, $0.05$ and $0.10$.
Each simulation is repeated 200 times.
The simulation results are given in Table \ref{tab:4a}.
We can see from Table \ref{tab:4a} that the power of the proposed test increases with the community size, network density $r$ and $z\%$, which characterizes the difference between $\sigma_0$ and the truth.
We can also see the proposed test is quite powerful against alternatives.
For example, in the case of $n/k=200$ and the network is sparse with $r=0.05$, the test can successfully reject the null when only $1\%$ labels in $\sigma$ are corrupted.
Under this setting, when $\sigma_0$ is moved away from the true $\sigma$, one can show that $(k_0,\sigma_0)\in\mathcal{F}(k,\sigma,B)$ and our test is powerful based on Theorem \ref{theorem:alternative:a}.

\end{document}